\documentclass[11pt]{article}
\usepackage{graphicx}
\usepackage{amsfonts, amssymb, amsmath}
\usepackage{booktabs}

\newtheorem{lemma}{{\bf \sc Lemma}}
\newtheorem{proposition}{{\bf \sc Proposition}}

\newtheorem{corollary}{{\bf \sc Corollary}}

\newtheorem{definition}{{\bf \sc Definition}}

\newcommand{\bge}{\begin{equation}}
\newcommand{\ene}{\end{equation}}

\def\eproof{\hbox{\hskip3pt\vrule width4pt height8pt depth1.5pt}}

\begin{document}

\title{Homophily and Long-Run Integration in Social Networks\thanks{
Following the suggestion of JET editors, this paper draws from two working
papers developed independently: \cite{BR10} and \cite{CJP10b}.
We gratefully acknowledge financial support from
the NSF under grant SES-0961481 and we thank Vincent Boucher for his
research assistance.
We thank the staff from the American Physical Society (APS) Journal Information Systems (http://publish.aps.org/departments/journalinformationsystems) for their assistance.
We also thank Habiba Djebbari, Andrea Galeotti, Sanjeev
Goyal, James Moody, Filippo Radicchi, Betsy Sinclair, Bruno Strulovici, and Adrien Vigier,  as
well as numerous seminar participants and the associate editor and anonymous referees for helpful
comments and suggestions.   } }
\author{Yann Bramoull\'e\thanks{%
Department of Economics, Universit\'e Laval and Aix-Marseille School of
Economics.} \and Sergio Currarini\thanks{%
Universit\`a di Venezia. Email: s.currarini@unive.it} \and Matthew O. Jackson%
\thanks{%
Department of Economics Stanford University, CIFAR, and the Santa Fe
Institute. Email: jacksonm@stanford.edu, http://www.stanford.edu/$\sim$%
jacksonm/} \and Paolo Pin\thanks{%
Dipartimento di Economia Politica e Statistica, Universit\`a degli Studi di
Siena (Italy). Email: paolo.pin@unisi.it} \and Brian W. Rogers\thanks{%
MEDS, Kellogg School of Management, and NICO, Northwestern University.}}
\date{\today}
\maketitle

\begin{abstract}
We model network formation when heterogeneous nodes enter sequentially and form
connections through both random meetings and network-based
search, but with type-dependent biases.  We show that there is ``long-run integration,'' whereby the composition of types
in sufficiently old nodes' neighborhoods approaches the global type distribution, provided
that the network-based search is unbiased. However, younger nodes'
connections still reflect the biased meetings process.
We derive the type-based degree distributions and group-level homophily
patterns when there are two types and location-based biases. Finally, we illustrate aspects of the model with an empirical
application to data on citations in physics journals.



\newpage{}
\end{abstract}

\section{Introduction}

Homophily patterns in networks have important consequences. For example,
citations across literatures can affect whether, and how quickly, ideas
developed in one field diffuse into another. Homophily also affects a
variety of behaviors and opportunities, with impact on the welfare of
individuals connected in social networks.\footnote{%
See \cite{J07,J08b,MSC01} for more background and discussion.}
In this paper we analyze a model that provides
new insights into the patterns and emergence of homophily, and we illustrate
its implications with an application to a network of scientific citations.

Our main objective is to study how homophily patterns behave in an evolving
network. Do nodes become more integrated or more segregated as they age? How
does this evolution depend on the link formation process? In particular,
does the network become more integrated if new connections are formed at
random or if they are formed \textit{through} the existing network?

To answer these questions, we study a stochastic model of network formation
in which nodes come in different types and types, in turn, affect the
formation of links. We accomplish this bye introducing individual heterogeneity
to the framework of Jackson and Rogers \cite{JR07}, allowing us to focus on the
issue of homophily generated through specific biases in link formation. A
new node is born at each time period and forms links with existing nodes.
The newborn node connects to older nodes in two ways. First, she meets nodes
according to a random, but potentially type-biased, process. Second, the
newborn node meets neighbors of the randomly met nodes (\textquotedblleft
friends of friends\textquotedblright ). This is referred to as the search
process and can also reflect type biases. To illustrate, consider citation patterns.
Typically when writing a new paper, some references are known or found by chance by the authors
while others are found \textit{because} they are cited in known papers. Biases arise because papers may cite references with greater frequency within their own
field. We examine the long-run properties of this model and the structure of
the emerging network.

The biases could arise from agents' preferences over the types of their
neighbors and/or from biased meeting opportunities that agents face in
connecting to each other. So, in one direction we enrich a growing network
model by allowing for types and biases in connections, and in another
direction we bypass explicit strategic considerations by studying a process
with exogenous behavioral rules. Since in the model search goes through
out-links only, strategic considerations are to some extent inherently
limited, since a node cannot directly increase the probability of being
found through its choice of out-neighborhood. While this may not be a good
assumption in some contexts, such as business partnerships or job contacts,
where search presumably goes both directions along a link, it is appropriate
in other contexts, such as scientific citations where the time order of
publications strictly determines the direction of search.

We wish to understand the conditions under which the network becomes
increasingly \textquotedblleft integrated\textquotedblright\ over time. We
consider three different notions of integration. Under \textit{weak
integration}, nodes who are old enough are more likely to get new links than
young nodes, independent of types. In this sense, age eventually overcomes
any bias in link probabilities. Under \textit{long-run integration}, the
distribution of types among the neighbors of a given node converges to the
population distribution as the node ages. This is a strong property
that requires biases among neighbors to eventually disappear, despite the
biased formation process. Finally, under\textit{\ partial integration}, the
type distribution among neighbors moves monotonically towards the population
distribution as nodes age, although it may maintain some bias in the limit.
These notions capture different, but related, aspects of the idea of network
integration.

Our main theoretical results are as follows. First, weak integration is
satisfied whenever the probability that a given node is found increases with
that node's degree. This holds in any version of our model where at least
some links are formed through search and there is some possibility of
connecting across types.%

In contrast, long-run integration holds only when search is unbiased. That
is, the random meeting process can incorporate arbitrary biases, but once
the initial nodes are met, the new node chooses uniformly from the set of
their neighbors, ignoring any further implication of their types. Finally,
we show that under a particular condition on the biases, the process evolves
monotonically and satisfies partial integration. In particular, the biases
in nodes'\ links generally decrease with age.


To understand where this tendency towards integration comes from, consider
unbiased search. Observe first that as a node ages, the proportion of his
links obtained through search approaches unity, since the number of
neighbors grows with age while the probability to be found at random
decreases with population size. Next, note that unbiased search does \textit{%
not }imply an absence of bias among the neighbors of randomly met nodes. Due
to homophily, randomly met nodes are relatively more connected with nodes of
their own types. A critical fact our analysis uncovers, however, is that
bias among neighbors' neighbors tends to be lower than among direct
neighbors. This is because some nodes of other types are found at random and
these nodes are relatively more connected to nodes like themselves. So the
set of neighbors'\ neighbors has a more neutral composition than the
neighborhoods of same-type nodes. Network-based search increases the
diversity of connections and, conversely, nodes found through search are
being found by a more diverse set of nodes. And since search plays a larger
role with age, older nodes are less biased in their connections.


In order to analyze network structure in more detail we consider a special,
but natural, two-type specification of the model where random meetings are
organized through a geographic or social space. Nodes of a given type are
more likely to reside in a given location and random meetings take place
without further bias in the various locations. In this model, biases in
random meetings are inherently tied in a precise way to the
type-distribution of the population. This feature allows us to obtain a
number of further results. In particular, we derive an explicit formula
relating a node's local homophily among neighbors to its age or degree. This
illustrates our general results and further shows how partial and long-run
integration are affected by changes in types'\ shares. We also study two
important structural properties of the resulting network that are less
tractable in the general model:\ degree distributions and group-level
homophily. We show how to modify the existing analysis of degree
distributions to account for individual heterogeneity and homophily,
obtaining new insights.

In addition, we obtain results on group-level homophily consistent with
empirical results presented in \cite{CJP09,CJP10a}. We
find that relative group size has an important impact on how meeting biases map
into aggregate properties of the network. In particular, relative homophily
in the network is strongest when groups have equal size, and vanishes as the
groups take increasingly unequal sizes. Turning to degree distributions, we
find that the majority and minority groups have different patterns of
interactions. In particular, for the minority group, links from their own
group are on the one hand rarer due to a size effect, but on the other hand
a homophilic bias pulls in the other direction, creating a tension in the
overall distribution of links. However a striking result that is, in
principle, testable is that the distribution of total in-links is identical
for the groups independent of their relative sizes.

Moving from the theoretical analysis, we illustrate the implications of the
model using data on scientific citations in journals of the American
Physical Society (APS) published between 1985 and 2003. We find that the
proportion of citations that a paper obtains from other papers in its own
field decreases as the paper ages and becomes more cited. 
The observed citation patterns provide some evidence of the partial integration property
and are at least partly consistent search follows a less biased (possibly unbiased) pattern in the citation process.
In studying this application we are motivated by two factors. First,
patterns of scientific citations have important welfare consequences as they
affect the diffusion of knowledge, with impacts on different research fields.%
\footnote{%
See, for instance, \cite{BL06,JT96}.}
Previous research, such as \cite{KNS10,PV04}, generalizing popular
concepts such as the \emph{recursive impact factor}, stress that the
importance of a citation relies on the paths that it allows in the network
of citations. We complement this argument by considering under which
conditions citations are likely to bridge scientific production across
different communities.\footnote{%
Ref. \cite{R_alii01} studies cross-field citations in the scientific
production of the 90's, for three different datasets.} Second, scientific
citations possess all the features of the network formation process that we
study: nodes (papers) appear in chronological order and never die, they link
directionally to previously born nodes, they have types (scientific
classifications), and they find citations both directly and though search
among the citations of other papers.\footnote{%
These longitudinal aspects of citation networks have motivated the use of
growing network models in previous papers including the seminal work on
citation networks by Price \cite{P65,P76}. Refs. \cite{Berner_alii04,SR07}, among others, find that citations on the
PNAS on a 20 years interval show some aspects of a bias towards recently
published papers, while Refs. \cite{N09,R98}, correcting for
cohort size and idiosyncratic popularity, find an age effect (first mover
advantage) and a frequency distribution of in-citations that are consistent
with a growing network model such as the one that we develop here. Finally,
Ref. \cite{STA09} find a positive correlation between homophily
of out-citations and the number of in--citations, but this effect is valid
only for low number of in-citations.}

More generally, our study contributes to a growing literature in economics
and other disciplines studying the causes and consequences of homophily in
social networks. Refs. \cite{CJP09,CJP10a} study a matching
process of friendship formation. They document several empirical patterns of
homophily and explain them through a combination of biases with respect to
choice and chance. By design, all individuals have the same degree and age
has little impact. In contrast, differences in age and degree are central to
our analysis. Ref. \cite{J08a} incorporates homophily into the random graph
model of \cite{CL02a,CL02b}.\footnote{%
Ref. \cite{GJ08} uses this extension to study how homophily affects
communication dynamics in networks, demonstrating explicitly one way in
which homophilic structure impacts outcomes, as does \cite{JL11}.} Again by design, homophily is
not affected by degree or age in this approach. Thus, our study and these
two papers study homophily patterns in networks from complementary
perspectives. In particular, we provide the first study of how homophily
patterns change over time and of the relation between homophily and a node's
degree.

This study also advances the analysis of stochastic models of network
formation. Earlier work has made great progress in explaining structural
network features such as small diameter, high clustering and fat tails in
degree distributions, see \cite{AlbertBarabasi99,CL02a,CL02b,JR07,N03,N04,StrogatzWatts98}.
However, most of these studies assume homogeneous agents and neglect
homophily. With respect to this literature,\ we develop and study one of the
first stochastic model of network formation incorporating individual
heterogeneity.

The rest of the paper is organized as follows. Section \ref{sec:model} presents the model
with bias only in the random meeting process. Section \ref{model_bias_only_random} includes the main
result about long-run integration in this setting. Section \ref%
{sec_location_bias} studies the special case of two-types and location based
bias. Section \ref{sec_model_biased_search} studies the integration
properties of the model when biases appear also in the search part of the
meeting process. Section \ref{sec_citations} contains the empirical
application to citation data.

\section{Homophily in a random meeting process}
\label{sec:model}

In our model, nodes are born with randomly assigned types and enter
sequentially, meeting existing nodes upon entry. Meetings result in
(directed) links. Meetings take place through two distinct processes, which
we refer to as \emph{random} and \emph{search}. The meeting processes depend
on the types of the nodes involved. In this section, we study the impact of
type-based biases on the random meeting process.

\subsection{The model}

\label{sec_intr_model}

Time is indexed by $t=1,2,....$. In each period a new node is born. We index
nodes by their birth dates, so that node $t$ is born in period $t$.

Nodes have ``types,'' with a generic type denoted $\theta$ belonging to a
finite set $\Theta$ (with cardinality $H$). A newborn's type is randomly
drawn according to the time invariant probability distribution $p$ (so that
types are i.i.d., across time).

A newborn node sends $n>1$ (directed) links to the nodes that were born in
previous periods. Of these $n$ links, a fraction $m_{r}$ selects nodes
according to a type-dependent \emph{random} process;\footnote{$m_{r}n$ is an
integer in the underlying process, but allowed to be arbitrary in the
mean-field continuous-time approximation we analyze.} these nodes are called
\textquotedblleft parents\textquotedblright . The remaining fraction $%
m_{s}=1-m_{r}$ selects nodes among the neighbors of the $nm_{r}$ parents
that have been found via the random process; we refer to this second part of
the process as \textquotedblleft search\textquotedblright .\footnote{%
In the underlying process, if some node is found to which the newborn is
already connected, then the node is redrawn. If there are too few new nodes
in the neighborhoods of the nodes found in the first part of the process,
then the random nodes are redrawn. To ensure that the process is
well-defined, we begin with a set of $n^{2}$ nodes in a sequence, each
connected to all predecessors.} We define $\sigma \equiv m_{r}/m_{s}$ to be
the ratio of the number of links formed by the random process to the number
of links formed by the search process.

Looking first at the random part of the process, we denote by $p(\theta
,\theta ^{\prime })$ the probability that a link sent by a node of type $%
\theta $ reaches a node of type $\theta ^{\prime }$. Among nodes of type $%
\theta ^{\prime }$, the link is formed uniformly at random, so there is no further discrimination in this part of the
process. If the random meeting process were unbiased, the probability $%
p(\theta ,\theta ^{\prime })$ would equal the share $p(\theta ^{\prime })$
of $\theta ^{\prime }$ agents in the system. When $p(\theta ,\theta ^{\prime
})\neq p(\theta ^{\prime })$ we say that there is bias. This can be
interpreted in different ways. One can view the bias as a reduced form for
preferences that nodes have over the type of connections they form. The case
of \textquotedblleft homophilistic\textquotedblright {}\ preferences for
type $\theta $ is then captured by a situation in which $p(\theta ,\theta
)>p(\theta )$. The bias could also arise from constraints in the meeting
process, or from spatial differentiation, as in the location model that we
will analyze in Section \ref{sec_location_bias}.\footnote{%
See \cite{CJP09,CJP10a,CJP10b} for more details on
other models that can justify this reduced form.}

Turning now to the search part of the process, the way in which friends are
drawn from parents' neighborhoods may be, in principle, either biased or
unbiased. Much of the paper will study a process with biases only in
the random part, so that in the search part, links are formed according to a
uniform distribution on the set of parents' neighbors. This assumption has
natural interpretations and various applications. It applies, for instance,
to cases where agents face a bias in meeting strangers, but then get to meet
the ``friends'' of their new friends without bias. When the original bias in
meetings comes from biased opportunities, this seems to be a natural
assumption; when the bias originates in preferences, it may still be the
case that this bias tends to vanish when meetings are mediated by friends.
In Section \ref{sec_location_bias} we will explicitly analyze a model that
relates these biases to location--based differences in the meeting process.
When search is itself biased, so that the additional $nm_{s}$ nodes are
found among parents' neighbors using a type-dependent probability
distribution, two types of biases are naturally defined: a bias that
discriminates according to the types of the parents through which search is
made, and a bias that discriminates according to the types of the parents'
neighbors. Which type of bias is more appropriate depends on the instance of
network one has in mind, and leads to formally different models of link
formation. In Section \ref{sec_model_biased_search} we study the case of
biased search and its consequences for integration.

Before formally deriving the dynamics of the various processes, in the next
section we propose three notions of integration that measure the extent to
which the bias in the random and/or search process translates into biases in
the long run type-patterns of link formation.

\subsection{Integration}

The definitions we provide capture different aspects of integration,
focusing on how a node's type-pattern of connections evolves with age, and
whether it gets progressively more (or less) integrated with the rest of the
network.

It is important to note that there are two different aspects of integration:
the evolution of newborns' newly formed links (out-links) and the evolution
of older nodes' incoming links. These will exhibit different dynamics. Given
the bias in the random part of the network formation process, it is clear
that there will always be some bias in the out-links of newborn nodes. The
main questions with regard to the out-links thus pertain to how the links
formed by search behave over time, and this is related to the question of
how the in-links of older nodes behave.

All three notions of integration discussed here pertain to the behavior of
in-degrees of nodes. Out-degree dynamics are studied in Sections \ref%
{sec_out_degree} and \ref{sec_formulas_agg}.

Our first notion requires, in particular, that old enough nodes are found by newborn nodes
with higher probability than younger nodes, independently of the types of
the nodes involved.

\begin{definition}
\label{weak}The network formation process satisfies the \textbf{weak
integration property} if for every $t_{0}$, there exists $t>t_{0}$ such
that, for all $t^{\prime}\geq t$ and for all $\theta\in\Theta$, the node
born at time $t^{\prime}$ has a lower probability than node ${t}_{0}$ to
receive a link from a node of type $\theta$ born at time $t^{\prime}+1$.
\end{definition}

Note that this form of integration requires that an old enough node of type $%
\theta$ ends up receiving a link from a newborn node of type $%
\theta^{\prime} $ with a higher probability than a young enough node of the
same type $\theta^{\prime}$ as the newborn. So, even if link formation
probabilities are biased in favor of similar nodes (homophily), old enough nodes
are found more often even when of a different type than the newborn.

This form of integration is rather weak, and does not bear implications on
the type-composition of any given node's in-degree. Our second notion of
integration requires that as nodes age, their local neighborhood grows to represent
more and more the type composition of the population. It is therefore a
``monotonicity'' property, requiring that integration, here defined in terms
of how close the composition of neighbors' types is to what would obtain in the unbiased case,
grows with age.

\begin{definition}
\label{partial} The network formation process satisfies the \textbf{partial
integration property} if for every node $t_{0}$ the fraction of each type $%
\theta$ in the in-degree of $t_{0}$ is weakly closer to $\theta$'s
population share at time $t^{\prime}$ than at time $t$, for $t^{\prime}>t$,
and strictly closer for some types.
\end{definition}

So, under partial integration, the in-neighbors of an agent become more and
more representative of the overall population as time elapses.

Our final notion of integration is stronger\footnote{%
By \emph{stronger} we do not mean that it is a necessary condition for the
partial integration property defined above. One could think of partial
integration as a criterion of monotonicity of a function in one variable,
while the \emph{long--run integration} defined below is a criterion of
convergence, that could however also be non--monotonic.} and requires that
nodes eventually attract in-links according to population shares.

\begin{definition}
\label{longrun} The network formation process satisfies the \textbf{long-run
integration property} if for every node $t_{0}$ the proportion of each type $%
\theta$ in the in-degree of $t_{0}$ converges to $\theta$'s population share
as node $t_{0}$ ages.
\end{definition}

In other words, in the long run any surviving difference in the proportion
of links received by old nodes from different types is due only to the
distribution of types in the population, and the biases in link-formation
have no consequences for eventual in-degree patterns.

\section{Integration with biased random meetings and unbiased search}
\label{model_bias_only_random}

\subsection{Model dynamics}

\label{sec_model_dynamic}

A benchmark model to study long run integration properties of link formation
is one where only the random part of the process is biased, and no further
bias is present in the search part of the process. More precisely, the
search process is unbiased in the sense that additional ties are found
through a uniform sample among parents' neighbors, but remains indirectly
biased through the bias that the random process has induced on the type
composition of the parents' neighborhoods. This model allows for a clear
understanding of the mechanics that lead to integration, and why and when
integration may fail.

We study a continuous time approximation of the model, using the techniques
of mean-field theory. This provides approximations and limiting expressions
of the process that ignore starting conditions and other short-term
fluctuations that can be important in shaping finite versions of the model,
and so the results must be viewed with the standard cautions that accompany
such approximations and limit analysis. We consider the expected change in
the discrete stochastic process as the deterministic differential of a
continuous time process.

Let us first look at the probability that node $j$ is found by newborn node $%
t+1$. This depends on the shape of the network that has formed up to time $t$%
. In particular, it depends on the type-profile of in-neighbors of $j$ at
time $t$, and on the bias of the newborn node towards such types. Since
search is not type-biased, each link that agent $t+1$ forms through search
is drawn from a uniform distribution over the set of all neighbors of all parent
nodes that agent $t+1$ has found at random.

Letting $P_{j}^{t}(\theta_{t},\theta_{j})$ denote the probability that a
node born in period $j$ of type $\theta_{j}$ receives a link from a node of
type $\theta_{t}$ born at time $t>j$, the following expression is a
mean-field approximation of the overall linking probability:
\begin{equation}
P_{j}^{t+1}(\theta,\theta_{j})=nm_{r}\frac{p(\theta)p(\theta,\theta_{j})}{%
tp(\theta_{j})}+nm_{s}\sum_{\theta^{\prime}\in\Theta}p(\theta)p(\theta,%
\theta^{\prime})\frac{\sum_{\lambda=j}^{t}P_{j}^{\lambda}(\theta^{\prime},%
\theta_{j})}{tp(\theta^{\prime})}\frac{1}{n}  \label{5}
\end{equation}

The first term on the right--hand side captures the probability of node $j$
being found at random. The probability that node $t+1$ is of type $\theta$
and links at random to a node of type $\theta_{j}$ is $p(\theta)p(\theta,%
\theta_{j})$. This is divided by the number of nodes of type $\theta_{j}$ at
time $t$ which, under a mean-field approximation, is equal to $%
tp(\theta_{j}) $. It is then multiplied by the number of links formed at
random, $nm_{r}$.

The second term is the probability of node $j$ being found through search.
It is given by the number of search links ($nm_{s}$) formed by the node born
at $t+1$, times the sum, over all possible types $\theta^{\prime}$, of the
probabilities that $j$ is found through a node of type $\theta^{\prime}$.
For each possible type $\theta^{\prime}$, this probability is given by the
joint probability of the following events (corresponding to the four terms
in the first summation over types): ($i$) the newborn node is of type $%
\theta $; ($ii$) it forms a link with a $\theta^{\prime}$-type node; ($iii$)
the $\theta^{\prime}$-type node has linked to $j$ since $j$ was born;%
\footnote{%
Note that this ratio has the total (expected) number of links received by
agent $j$ from $\theta^{\prime}$ agents up to time $t$ as numerator, and the
total number of $\theta^{\prime}$ nodes in the system at time $t$ as
denominator} ($iv$) among the $n$ neighbors of this $\theta^{\prime}$-type
node, that exactly $j$ is found.

It is useful to express the terms of the above formula in a compact way. For
all $\theta$, $\theta^{\prime}$ we write
\begin{equation*}
B_{r}(\theta,\theta^{\prime})\equiv p(\theta)\frac{p(\theta,\theta^{\prime})%
}{p(\theta^{\prime})}.
\end{equation*}
Note that the ratio $\frac{p(\theta,\theta^{\prime})}{p(\theta^{\prime})}$
in the above expression is a measure of the bias that type $\theta$ applies
to type $\theta^{\prime}$, so that when this ratio is $1$ there is no bias,
while when it is greater (less) than $1$ there is a positive (negative) bias
of type $\theta$ towards type $\theta^{\prime}$. In the case of no bias, $%
B_{r}(\theta,\theta^{\prime})$ is simply the probability of birth of a type $%
\theta$ node, and $P_{j}^{t+1}(\theta,\theta_{j})$ is $n$ times the joint
probability that the newborn node is of type $\theta$ and that node $j$ is
found by drawing uniformly at random from a population of $t$ nodes. We can
decompose the matrix $\mathbf{B}_{r}$ as the product of two matrices $%
\mathbf{A}$ and $\mathbf{Q}$, where $\mathbf{A}$ may be seen as a transition
matrix of a Markov process (a Markov matrix),\footnote{%
In \ref{matrices} we derive some general results on Markov matrices that
will be useful in \ref{proofs}, where we prove our propositions.} and $%
\mathbf{Q}$ is a diagonal matrix where the diagonal is a probability vector:
\begin{equation}
\mathbf{B}_{r}=\mathbf{QAQ}^{-1},  \label{15}
\end{equation}
with
\begin{equation*}
\mathbf{A}_{\theta\theta^{\prime}}=p(\theta,\theta^{\prime})\ \ \ \ %
\mbox{and}\ \ \ \ \mathbf{Q}=\left(%
\begin{array}{ccc}
p(1) & ... & 0 \\
... & ... & ... \\
0 & ... & p(H)%
\end{array}%
\right)\ \ .
\end{equation*}
Using the matrix $\mathbf{B}_{r}$, equation (\ref{5}) becomes:
\begin{equation}
\mathbf{P}_{t_{0}}^{t+1}=\frac{nm_{r}}{t}\mathbf{B}_{r}+\frac{m_{s}}{t}%
\mathbf{B}_{r}\sum_{\lambda=t_{0}}^{t}\mathbf{P}_{t_{0}}^{\lambda}\ \ ,
\label{6}
\end{equation}
where $\sum_{\lambda=t_0}^{t}\mathbf{P}_{t_{0}}^{\lambda}$ expresses the
expected in-degree, type by type, after time $t$ for a node born at time $t_{0}$%
.

We define
\begin{equation*}
\Pi_{t_{0}}^{t}\equiv\sum_{\lambda=t_{0}}^{t}\mathbf{P}_{t_{0}}^{\lambda}.
\end{equation*}
With a continuous approximation:
\begin{equation*}
\frac{\partial}{\partial t}\Pi_{t_{0}}^{t}=\mathbf{P}_{t_{0}}^{t+1}.
\end{equation*}
We study equation (\ref{6}) in terms of ordinary differential equations in
matrix form:
\begin{eqnarray}
\frac{\partial}{\partial t}\Pi_{t_{0}}^{t} & = & \frac{nm_{r}}{t}\mathbf{B}%
_{r}+\frac{m_{s}}{t}\mathbf{B}_{r}\Pi_{t_{0}}^{t},  \label{10}
\end{eqnarray}
with the initial condition
\begin{equation*}
\Pi_{t_{0}}^{t_{0}}=\mathbf{0.}
\end{equation*}

From now on we will always assume that $\mathbf{B}_{r}$ is invertible (so
that the specification of types is not redundant). With this assumption, the
unique solutions to these differential equations are the following:
\begin{eqnarray}
\Pi_{t_{0}}^{t} & = & n\frac{m_{r}}{m_{s}}\left(\left(\frac{t}{t_{0}}%
\right)^{m_{s}\mathbf{B}_{r}}-\mathbf{I}\right),  \label{13}
\end{eqnarray}
where a constant to the power of a matrix is defined as follows:
\begin{equation}
\left(\frac{t}{t_{0}}\right)^{m_{s}\mathbf{B}_{r}}=e^{\left(m_{s}\log\left(%
\frac{t}{t_{0}}\right)\mathbf{B}_{r}\right)}=\sum_{\mu=0}^{\infty}\frac{%
\left(m_{s}\log\left(\frac{t}{t_{0}}\right)\mathbf{B}_{r}\right)^{\mu}}{\mu!}%
.  \label{14}
\end{equation}

\subsection{Integration}

\label{sec_no_bias_search}

Let us test the various notions of integration on this model with unbiased
search.

It is clear that the model with $m_{r}=1$ cannot satisfy weak integration.
We show instead that whenever there is some degree of search ($m_{r}<1$)
weak integration is satisfied. In fact, in Section \ref%
{sec_model_biased_search} we strengthen this result to show that weak
integration is still satisfied when search is biased as well.

\begin{proposition}
\label{prop_1} If $m_{r}<1$, the model with unbiased search satisfies the
\emph{weak integration property}.
\end{proposition}

The proof (which appears, along with the proofs of our other results, in \ref{proofs}) shows that the weak integration property is not
specific to the unbiased search model. Indeed, various models in which the
in-degree of a node determines the probability of being found by a newborn
node in a sufficiently increasing manner would give the same result.
Moreover, search is not needed for this type of dependence to take place.
Another model with \textquotedblleft type-biased\textquotedblright {}\
preferential attachment in which the probability of receiving a link is
positively correlated with a node's in-degree, and which exhibits the same
weak integration property, is discussed in the conclusion of \cite{CJP10b}.

Partial and long-run integration are, again, not satisfied when $m_{r}=1$.
The next propositions show that, otherwise, the long--run integration property is always
satisfied by the model, while the partial integration property needs an
additional assumption.

\begin{proposition}
If $m_{r}<1$, the model with unbiased search satisfies the \emph{long--run
integration property}. \label{prop_lri}
\end{proposition}

Partial integration, instead, occurs under an additional condition. Consider
a Markov matrix $\mathbf{M}$. As formally stated in \ref{matrices}, writing $\bar{\mathbf{M}}\equiv\lim_{\mu\rightarrow\infty}\mathbf{M}^{\mu}$, we
say that $\mathbf{M}$ satisfies the \emph{monotone convergence property} if,
for every pair $i,j\in\{1,\dots,H\}$, and for every $\mu\in\mathbb{N}$, the
element $M_{ij}^{\mu}$ satisfies:

\begin{enumerate}
\item if $M_{ij}>\bar{M}_{ij}$, then $M_{ij}\geq M_{ij}^{\mu}\geq
M_{ij}^{\mu+1}\geq\bar{M}_{ij}$;

\item if $M_{ij}<\bar{M}_{ij}$, then $M_{ij}\leq M_{ij}^{\mu}\leq
M_{ij}^{\mu+1}\leq\bar{M}_{ij}$.
\end{enumerate}

The monotone convergence property captures the idea that transition
probabilities are monotone over time. Even with a strictly positive
transition matrix, this condition does impose additional restrictions.%
\footnote{%
As a simple illustrating example, consider a Markov process with three
states where transitions from state 1 to state 2, 2 to 3, and 3 to 1 occur
with high probability, and with the other transitions occurring with small
but positive probabilities. Then in one period going from 1 to 2 is likely,
but then it is unlikely to occur in two periods or three periods, but more
likely in four periods, and so forth. Things eventually converge to equal
likelihood on all states, but convergence is not monotone. One can also find
such examples that are more complicated where homophily is present.} It is
beyond the scope of this paper to find general or even necessary conditions
for monotone convergence of Markov matrices.

We then have the following result.

\begin{proposition}
If $m_{r}<1$ and $\mathbf{A}$ satisfies the monotone convergence property,
then the model with unbiased search satisfies the \emph{partial integration
property}. \label{prop_partial_no_bias}
\end{proposition}

Let us focus on the intuition behind the long run integration property of
the model with unbiased search. To fix ideas, let us examine the case in
which random probabilities have a homophilous bias. A given node can be
found by a newborn node of a different type via search in different ways:
one is that the newborn finds a neighbor of the given node that is of the
same type as the newborn, and another is that the newborn finds a neighbor
of the given node that is of the same type as the given node. The first way
is relatively more likely given the homophilous bias in the random part of
link formation, but the fact that this can also occur via the second route
leads this process to be less biased over time. Once the process has become
less biased, it even easier to be found by nodes of other types, and so the
neighborhood becomes even less biased, and this trend reinforces itself
leading to an unbiased process in the limit. To summarize, as a node ages it
becomes more of a ``hub'', attracting many links from all
types in the search process. This property, that also underlies the weak
integration property, together with unbiased search further decreases the
bias in the in-degree of hubs. As a result, the type composition of new
connections becomes even less biased for these hubs, and eventually the bias
is eliminated.

The way in which an individual's neighborhood composition limits to the
population frequencies as it ages is non-trivial. Notice that if a particular
individual became connected to a large proportion of others over time, then
his neighborhood would necessarily approximate population frequencies.
However, we emphasize that in our model, even though an individual's degree
grows without bound, the proportion of others to whom he is connected still
vanishes over time, so this effect is not what drives integration. This
happens because the entry rate of new individuals is constant, while the
probability for existing individuals to acquire a new link in any given period
goes to zero.

Finally, we remark that, while the neighorhood of every node approaches a
composition that reflects the aggregate population frequencies, it converges
to that distribution from a point that is affected by biases in the link
formation process. Since those links are perpetually being formed and are
subject to biases, the system never approaches a network that has unbiased
link patterns. Rather, it is always the oldest nodes in the system that have
the least biased neighborhoods. In fact, one way to see the persistent bias
is to focus on out-degree rather than in-degree. Thus, we turn now to
analyzing links by tracking where they originate.

\subsection{On the dynamics of out-degrees}
\label{sec_out_degree}

So far we have mostly focused on the dynamics of agents' in-degree. Of
course, out- and in-degree dynamics are intimately related, as the search
part of young nodes' out-degree will consist predominantly of old nodes,
with respect to whom the search part of the process is both directly and
indirectly unbiased (see Section \ref{sec_formulas_agg}). Here we take a
close look at the composition of out-degrees and how they evolve over time.
This is of interest not only to better understand integration, but also to
shed light on the evolution of homophily, that is, the tendency to form ties
with agents of the same type.

We first look at the steady state composition of the out-degree. Let us
denote by $d_{ij,t}$ the proportion of links that originate from a node of
type $i$ born at time $t$ that are directed towards nodes of type $j$. The
evolution of these proportions is given by:

\begin{equation}
d_{ij,t+1}=(1-m_{s})B_{r}(i,j)+m_{s}\sum_{h=1}^{H}B_{r}(i,h)\frac{%
\sum_{\tau=1}^{t}d_{hj,\tau}}{t}.
\end{equation}

The out--degree depends on the random part (first term) and on the search
part (second term) through the average out--degree of existing nodes. In
matrix form, this is written as follows:

\begin{equation}
\mathbf{D}_{t+1}=(1-m_{s})\mathbf{B}_{r}+m_{s}\mathbf{B}_{r}\frac{%
\sum_{\tau=1}^{t}\mathbf{D}_{t}}{t}.  \label{out_degree_diff_eq}
\end{equation}

To get a feeling for the limit of this process, it is useful to examine the
steady state $\mathbf{{\bar{D}}}$ of this system. The steady-state is such
that the out-degree of each type remains unchanged in time:

\begin{equation}
\mathbf{\bar{D}}=(1-m_{s})\mathbf{B}_{r}+m_{s}\mathbf{B}_{r}\mathbf{\bar{D}}%
\ \ .  \label{steady_state_outdegree}
\end{equation}

\begin{proposition}
\label{prop_outdegree} If $m_{s}<1$, then the steady state equation (\ref%
{steady_state_outdegree}) has a unique solution $\mathbf{\bar{D}}$, and the
system in (\ref{out_degree_diff_eq}) converges to $\mathbf{\bar{D}}$. %
\hbox{\hskip3pt\vrule width4pt height8pt depth1.5pt}
\end{proposition}

For $m_{s}<1$, the second term approaches the null matrix as $%
t\rightarrow\infty$. As long as the matrix $\mathbf{D}_{1}$ is more biased
than the steady state $\bar{\mathbf{D}}$ (which is true for $\mathbf{D}_{1}$=%
$\mathbf{A}$),
the bias in excess of the steady state decreases with time, vanishing in the
long run (see also \ref{matrices}).

This means that the biases in the out-links formed by agents decrease over
time, consistent with the homogenization of the search process and the
in-degree of older nodes which are dominating the search part of the
process. However, unlike the case of the in-degree of old nodes, full
homogenization does not occur even in the limit, since the random part of
the out-degree formation does not vanish over time.

\section{Location--based biases}

\label{sec_location_bias}

In this section we consider a specific form of bias in random meetings, and
restrict the analysis to two types for simplicity. By making explicit how
the bias in random meetings is generated, we accomplish two goals. First, we
generate a closed form expression that describes the integration of
individuals as they age. This formula allows us to study in more detail the
integration process, and provides parameters which can be empirically
estimated. Second, we obtain additional results on other features of the
network, specifically on aggregate homophily at the group level and
in-degree distributions that are type-sensitive. For each of these
categories of results, the location-based nature of the meeting biases
allows us to study the impact of changes in population frequencies on the
structure and properties of the emerging networks.

Nodes belong to one of two types: $\theta_{1}$ and $\theta_{2}$. With an
abuse of notation, wee let $p(\theta_{1})=p$ and $p(\theta_{2})=1-p$ in this
section. There are two locations $L^{1}$ and $L^{2}$. All biases in the
meeting process are captured by the parameter $\gamma\in\lbrack1/2,1]$,
which represents the probability that a $\theta_{i}$ node goes to location $%
L^{i}$, $i=1,2$.\footnote{%
The analysis below assumes away some implicit correlations in the meeting
process described here. This is as if (modulo matching issues) each new node
in $g^{i}$ spends a proportion $\gamma$ of his time in $L^{i}$, and the
probability of meeting any existing node is proportional to the time spent
with it in the same location.} Once assigned to a location, each agent meets
$m_{r}n$ nodes uniformly at random among all individuals present at this
location. Thus, it is simply the resulting composition of types in the two
locations that permits any type-dependent biases in the model. We maintain the assumption that
the search part of the process is unbiased.

At any time $t$, the expected number of $\theta_{1}$ nodes at $L^{1}$ is $%
p\gamma t$, while the expected number of $\theta_{2}$ nodes in $L^{1}$ is $%
(1-p)(1-\gamma)t$. Thus, the proportion of $\theta_{1}$ nodes in $L^{1}$ is $%
\frac{p\gamma}{p\gamma+(1-p)(1-\gamma)}$ while the proportion of $\theta_{1}$
nodes in $L^{2}$ is $\frac{p(1-\gamma)}{p(1-\gamma)+(1-p)\gamma}$. The
probability that a node of type $\theta_{1}$ links at random to a node of
the same type is thus:
\begin{equation}
p(\theta_{1},\theta_{1})=\gamma\frac{p\gamma}{p\gamma+(1-p)(1-\gamma)}%
+(1-\gamma)\frac{p(1-\gamma)}{p(1-\gamma)+(1-p)\gamma}  \label{b1 gM}
\end{equation}
and $p(\theta_{2},\theta_{2})$ is obtained by symmetry exchanging $p$ and $%
1-p$.

Thus, the model generates a simple explicit relation between population
frequencies and random meeting biases, controlled by the parameter $\gamma$.
Note that when $\gamma=1/2$, locations are independent of types and there is
no bias:\ $p(\theta_{i},\theta_{i})=p(\theta_{i})$. In contrast, when $%
\gamma>\frac{1}{2}$ random meetings are biased towards own group and $%
p(\theta_{i},\theta_{i})>p(\theta_{i})$. At the extreme when $\gamma=1$,
locations are perfectly correlated with types and individuals meet others
only from their own group so that $p(\theta_{i},\theta_{i})=1$. This allows
us to derive a number of comparative statics results with respect to
population frequencies below.

\subsection{Explicit formulas for long--run integration}

\label{sec_formulas_integ}

Using the expressions above we have
\begin{equation}
B_{r}=\left(%
\begin{array}{cc}
p(\theta_{1},\theta_{1}) & \frac{p}{1-p}(1-p(\theta_{1},\theta_{1})) \\
\frac{1-p}{p}(1-p(\theta_{2},\theta_{2})) & p(\theta_{2},\theta_{2})%
\end{array}%
\right).
\end{equation}

We note that Proposition \ref{prop_partial_no_bias} (together with Lemma \ref%
{matrix_2x2} in \ref{matrices}) implies that if $m_{r}<1$, then the
location--based model always satisfies the \emph{partial integration property%
}, because $p(\theta_{1},\theta_{1})\geq\frac{1}{2}$ and $%
p(\theta_{2},\theta_{2})\geq\frac{1}{2}$.

We can now solve equation (\ref{10}) in terms of $p$ and $\gamma$ to obtain

\begin{lemma}
The in-link composition at time $t$ of a type $\theta_{1}$ node born at time
$t_{0}$ is
\begin{eqnarray}
\Pi_{t_{0}}^{t}(1,1) & = & n\frac{m_{r}}{m_{s}}\left(p(\frac{t}{t_{0}}%
)^{m_{s}}+(1-p)(\frac{t}{t_{0}})^{bm_{s}}-1\right)  \label{degGrowthI} \\
\Pi_{t_{0}}^{t}(2,1) & = & n\frac{m_{r}}{m_{s}}(1-p)\left((\frac{t}{t_{0}}%
)^{m_{s}}-(\frac{t}{t_{0}})^{bm_{s}}\right),  \label{degGrowthII}
\end{eqnarray}
with the analogous expressions for type $\theta_{2}$, where
\begin{equation*}
b\equiv p(\theta_{1},\theta_{1})+p(\theta_{2},\theta_{2})-1=1-\frac{%
\gamma(1-\gamma)}{p(1-p)(2\gamma-1)^{2}+\gamma(1-\gamma)}.
\end{equation*}
\label{L1}
\end{lemma}

We can now show that the location model generates a simple, explicit
relationship between integration and in--degree at the individual level.
This allows us to illustrate the results from the previous section and to
obtain further predictions on the shape of integration. In particular, we
find that the amplitude of integration tends to be lower in larger groups.

We denote an individual's in--degree by $k$, which is a function of the
node's entry date $t_{0}$ and the individual's age. Then $r^{j}(k)$ denotes
the individual share of same type in--links for a node of type $\theta_{j}$
at the time when its degree is $k$.


\begin{proposition}
\label{Prop3} 
Suppose that biases are location-driven. Then, $r^{j}(k)$ is described by
\begin{equation*}
r^{j}(k)=p(\theta_{j})+(1-p(\theta_{j}))\frac{(1+km_{s}/(nm_{r}))^{b}-1}{%
km_{s}/(nm_{r})}.
\end{equation*}
The individual share of same type in--links is thus decreasing with $k$ and
convex in $k$, converging asymptotically to the population shares. Moreover,
for $k>0$, $\partial r^{j}/\partial p^{j}(k)>0$ and $\partial^{2}r^{j}/%
\partial p^{j}\partial k(k)>0$.
\end{proposition}

Thus, consistent with the general long-run integration result of Proposition %
\ref{prop_lri}, $r^{j}(k)$ converges to the population frequency $p(\theta
_{j})$ as $k$ and, hence, time, diverge. The convergence is monotonic, and
so satisfies the property of partial-integration described in Proposition 3.
Notice that by application of Lemma D in Appendix A, one could demonstrate
partial integration without the explicit solution contained in Proposition %
\ref{Prop3}. However, the formula delivered by Proposition \ref{Prop3}
allows us to derive some further implications. First, the relation between $%
r^{j}(k)$ and in--degree is convex, so its decrease with age takes place at
a decreasing rate. Second, and perhaps more importantly, the relation
between $r^{j}(k)$ and degree tends to be flatter in larger groups; the
difference in integration between low-degree and high-degree nodes is
smaller in larger groups. Third, this function could be readily fitted to
data. In contexts where information on how links are formed is lacking, this
could provide the basis of an empirical analysis of the model. This approach
is illustrated in \cite{BR10}.

\subsection{Cumulative link distributions}

\label{sec_cumulative_links}

We turn now to a more detailed discussion of the distribution of links
across nodes in the network. Proposition \ref{Prop3} makes explicit the
relationship between the degree of an individual and the local composition
of its in--neighbors, demonstrating, in particular, the properties of
partial and long-run integration. Integrating that relationship in order to
obtain a measure of group-level homophily requires knowing the distribution
of in-degree across individuals. This section is concerned precisely with
analyzing those degree distributions, which have become a cornerstone of
social network analysis.

Even with only two groups, capturing the distribution of links becomes
substantially more complex, relative to the one-group case, as nodes can
connect to both same- and different-type nodes, and one wants to keep track
of the different kinds and sources of links. In this context, we can keep
track of \textit{seven} different degree distributions rather than one.
Define $F_{ij}$ as the distribution of the in-degrees of type $\theta_{i}$
nodes paying attention only to links coming from nodes of type $\theta_{j}$,
$i,j=1,2$. Then $F_{1}$ and $F_{2}$ are the standard in-degree distributions
of $\theta_{1}$ and $\theta_{2}$ nodes (ignoring the types of neighbors),
and finally $F$ is the total in-degree distribution of the entire society.

We observe that as a consequence of Lemma \ref{L1}, all of the degree
distributions have a power-law upper tail, as has been documented
extensively in empirical contexts, starting from \cite{AlbertBarabasi99}.
Further, we are able to make predictions about how the distribution of links
from own- and other-group nodes relate to each other, making clear the
importance of whether a node is in the majority or minority group. Finally,
we show that changing the bias in location-based meetings causes the
distributions to shift in the sense of first order stochastic dominance
(Proposition \ref{Prop3}).

To begin the analysis consider, for example, $F_{1}$.  Observe that $%
F_{1}(k)=1-t_{0}/t$, where $t$ is an arbitrary time period and, by definition, $t_{0}$ is
the node that has in-degree $k$ at time $t$. $t_{0}$ can be solved for under
the mean-field approximation. This defines $F_{1}$ implicitly as a function
of $k$, and an analogous method works for the other distributions. While
these equations do not usually yield closed-form solutions, they still allow
us to derive important properties of the degree distributions. Our first
such result orders the degree distributions as the number of out-links is
varied.

We first ask how $F_{i1}$ and $F_{i2}$ compare to each other. That is, we
focus on one group $\theta_{i}$, and for those nodes we compare the
(distributions of) the number of links coming from the the own group and the
other group. We find that the relationship depends on the size of
group $i$.

\begin{proposition}
\label{3part prop} Fix $p>1/2$ so that the majority group is group 1. Then

(i) $F_{11}$ FOSD $F_{12}$;

(ii) If $\gamma<1$, then $F_{22}$ never FOSD $F_{21}$;

(iii) $F_{21}$ FOSD $F_{22}$ if and only if $b\leq\frac{2p-1}{2p}$;

(iv) $F_{1}=F_{2}$
\end{proposition}

These results express the interplay of two effects. On the one hand, there
is a direct size effect through which nodes receive more links from the
larger group. On the other hand, homophily leads nodes to receive relatively
more links from nodes of the same group. In the larger group, both effects
are aligned which implies that $F_{11}$ FOSD $F_{12}$. In the smaller group,
however, these effects pull in opposite directions.

The third item in the proposition says that if homophily is not too large,
the size effect dominates and $F_{21}$ FOSD $F_{22}$. The condition in part
(iii) requires that $b$, and hence $\gamma$, be lower than or equal to some
threshold value.\footnote{%
Using the definition of $b$, from Lemma \ref{L1}, the threshold can be
written $\gamma(1-\gamma)\geq p(1-p)/(1+2(1-p)(2p-1))$.} We can see that
this threshold is increasing in $p$. As the size of the larger group
increases, the size effect becomes relatively more important and $F_{21}$
ends up dominating $F_{22}$ for a larger range of the parameters. In
contrast, the second item says that even if homophily is large, as long
as it is not perfect ($\gamma<1$), the homophily effect cannot dominate the
size effect. The explanation lies with nodes of high degree. We can show
that, in the upper tail, $F_{22}$ always lies above $F_{21}$. In other words, the
size effect dominates for the hubs of the smaller group, and they tend to
get relatively more connections from nodes of the larger group. This is
related to partial integration: the largest degree nodes are more
integrated with respect to their in-degree. In other words, the hubs in the minority group have
the greatest proportion of their in-neighbors from the majority group.

The last part provides a particular empirical prediction, as
independent of the homophily biases, the relative group sizes, and the
proportion of links formed through the random meeting process, the in-degree
distributions of the two groups must be identical.

The final result in this section describes how the distributions of inter-
and intra-group links respond to changes in the meeting bias.

\begin{proposition}
\label{Prop_fosd} Assume biases are location-driven. Fix $p,m$ and $\sigma$
and take $\gamma<\gamma^{\prime }$. Let $F_{ij}$ be the distributions
corresponding to $\gamma$ and let $F^{\prime }_{ij}$ be the distributions
corresponding to $\gamma^{\prime }$, for $i,j=1,2$. Then $F^{\prime }_{11}$
and $F^{\prime }_{22}$ strictly FOSD $F_{11}$ and $F_{22}$, while $F_{21}$
and $F_{12}$ strictly FOSD $F^{\prime }_{21}$ and $F^{\prime }_{12}$.
\end{proposition}

When the meeting bias increases, no matter the group sizes, individuals tend
to form more links within their own groups, and fewer links across groups.

\subsection{Long run homophily and group size}

\label{sec_formulas_agg}

This section complements the general analysis of Section \ref{sec_out_degree}
with a detailed inspection of the steady state out-degree composition in the
two type location based model. In this specific context, our aim is to
understand how the built-in homophily in random meetings translates into biased
long run proportions of out-links that stay within a group, and how these
proportions relate to groups' frequencies. Proposition \ref{Prop5} below
tells us that group-level homophily is strongest when the groups have nearly
equal sizes, and vanishes at the extreme when one group dominates society.
Further, Corollary \ref{Coroll_shift} tells us that, for given population
frequencies, group-level homophily is stronger when the bias in random
meetings or the relative prevalence of random meetings is higher.

In preparation for this result, define the homophily index $H(\theta_{i})$
as the expected proportion of the links formed by a new $\theta_{i}$ node
that are with same-type nodes. At the steady state, $H(\theta_{1})$ and $%
H(\theta_{2})$ satisfy the following equation
\begin{equation*}
H(\theta_{1})n=m_{r}np(\theta_{1},\theta_{1})+m_{s}n[H(\theta_{1})p(%
\theta_{1},\theta_{1})+(1-H(\theta_{2}))(1-p(\theta_{2},\theta_{2})],
\end{equation*}
as well as its symmetric counterpart. We know from the results of Section
\ref{sec_out_degree} that the steady state solution will be greater than $p(\theta_1)$, as
the bias in out-links does not vanish with time, since the random meeting
process always constitutes a non-trivial portion of out-degree. This
equation decomposes the expected number of links formed within group as the sum of
two terms capturing links formed at random and links formed through search.
At random, this number is by definition proportional to the probability $%
p(\theta_{1},\theta_{1})$. Through search, there are two ways to connect
within the group, depending on whether the intermediary node is of the same
type or not. Solving these equations yields, for type $\theta_{1}$,
\begin{equation*}
H(\theta_{1})=\frac{p(\theta_{1},\theta_{1})\sigma+1-p(\theta_{2},\theta_{2})%
}{\sigma+2-p(\theta_{1},\theta_{1})-p(\theta_{2},\theta_{2})},
\end{equation*}
recalling that $\sigma=m_{r}/m_{s}$ represents the ratio of the number of
links formed at random to the number of links formed through search.
Combining this expression with equation (\ref{b1 gM}), we obtain an explicit
formula linking group homophily and population frequencies, controlled by
the ratio of random to search meetings $\sigma$ and the bias parameter $%
\gamma$. In particular, we can see that $H(\theta_{i})$ is increasing in $%
\sigma$. As $\sigma$ tends to zero and search dominates the network's
formation, $H(\theta_{i})$ tends to $p(\theta_{i})$, while when $\sigma$
becomes large and random meetings dominate, $H(\theta_{i})$ tends to $%
p(\theta_{i},\theta_{i})$.

This means that the larger the role of search in the network formation
process, the more integrated the society becomes, consistent with the
intuition obtained from the general setting. This provides an expression at
the group-level of the idea that search tends to reduce the imposed bias in
meetings. The reason for this is that nodes found through search are more
likely to be of the other type than nodes found at random, due to the
possibility that the intermediary node is of the other type. As search
dominates, homophily disappears completely and links are formed according to
population frequencies. On the other hand, when the random meeting process
dominates, links are formed according to the probabilities determined by the
location-based meetings.

To analyze how group-level homophily varies with group size, we find it
useful to look at a normalized index introduced by \cite{Coleman58} (see \cite{CJP09}). Define \textit{relative homophily } (or
\emph{imbreeding homophily}) as follows
\begin{equation*}
IH(\theta_{i})=\frac{H(\theta_{i})-p(\theta_{i})}{1-p(\theta_{i})}\ \ .
\end{equation*}
Relative homophily is positive when a group forms a higher proportion of its
links within the group than would be implied by the population sizes, and is
normalized to have a maximal value at unity. Again, from the results of
Section \ref{sec_out_degree}, relative homophily will be positive in steady state. However,
we can now demonstrate a more detailed relationship between relative
homophily and relative group size. The following result shows how relative
homophily changes as the composition of society varies.

\begin{proposition}
\label{Prop5} $IH_{1}$ is symmetric around $p=1/2$. It is equal to zero at $%
p=0$ and $1$; it increases from $p=0$ to $p=1/2$, and decreases from $p=1/2$
to $p=1$, and is concave.
\end{proposition}

Thus, in the extreme cases where one group dominates society, relative
homophily disappears. Natural mixing occurring at each location tends to
homogenize meetings, and this effect overcomes the impact of location biases
when sizes are asymmetric. In all other cases, however, relative
homophily is positive, and is strongest for intermediate size groups,
reaching a maximum when the groups have equal size. Interestingly, the
equilibrium mixing model of \cite{CJP09} generates an
analogous result through a different analysis, and this prediction is
supported empirically looking at racial composition in the AddHealth data.

The next result shows how relative homophily responds to changes in the
meeting process.

\begin{corollary}
\label{Coroll_shift} $IH_{1}(p)$ is shifted up by an increase in $\sigma$ or
by an increase in $\gamma$.
\end{corollary}

That is, for a given society and homophily biases, decreasing the role of
search-based meetings increases relative homophily. This results from the
``dampening effect\textquotedblright{}\ of search-based meetings on network
homophily. The more prevalent are search-based meetings in the formation
process, the lower homophily will be, as ``friends of friends%
\textquotedblright{}\ are less likely to be of the same type than are
individuals met through the biased random meetings. Finally, increasing the
location bias, all else equal, increases relative homophily for any value of
$\sigma$, since this is the parameter that controls the extent to which
random meetings are exogenously biased.

\section{Integration with biased search}

\label{sec_model_biased_search}

We now allow for the search part of the process to be
biased as well as the random part. We do this for two main reasons. First,
we want to assess what degree of integration is still compatible with this
more flexible specification of biases in the network formation process. Second, the more realistic
assumption of some form of bias also in the search process is needed to
match the empirical patterns of scientific citations that we study in the
final section of this paper.

To prepare for this more general version of the model, remember that in the
analysis of Section \ref{model_bias_only_random}, the random bias affects
the choice of parents that are used to find the additional $nm_{s}$ search
connections. In general, the bias that affects the search process may differ
from, and possibly be a reinforcement of, the bias induced by the random
process. This additional bias is described via an $H\times H$ matrix where
each element is positive and of the form $B_{s}(\theta ,\theta ^{\prime })$.%
\footnote{%
There are constraints on this bias matrix to have the resulting output be
well-defined probabilities, but much can be deduced for general forms of the
matrix, and so we only specify the (obvious) constraints as they become
necessary. A more rigorous treatment and some explicit examples are in
\cite{CJP10b}.} A value of $1$ indicates no additional
bias, while a value greater (less) than $1$ indicates a positive (negative)
search bias of type $\theta $ towards type $\theta ^{\prime }$. This is
the distortion in the relative probabilities with which type $\theta $ searches
the out-neighborhoods of its parent nodes.

The mean-field approximation of the process is described by
\begin{equation}
\begin{array}{rcl}
P_{j}^{t+1}(\theta,\theta_{j}) & = & \frac{nm_{r}}{t}B_{r}(\theta,%
\theta_{j})+\frac{m_{s}}{t}\sum_{\theta^{\prime}=1}^{H}B_{r}(\theta,\theta^{%
\prime})B_{s}(\theta,\theta^{\prime})\sum_{\lambda=j}^{t}P_{j}^{\lambda}(%
\theta^{\prime},\theta_{j})\ \ .%
\end{array}
\label{rtb3}
\end{equation}

The product $B_{r}(\theta,\theta^{\prime})B\mathbf{_{s}}(\theta,\theta^{%
\prime})$ in (\ref{rtb3}) describes the probability applied by type $\theta$
to the selection of random nodes in search of type $\theta_{j}$. Note that
the bias is independent of both time variables $j$ and $t$, and of the type $%
\theta_{j}$ of the target.

In matrix form, the system becomes:
\begin{equation}
\mathbf{P}_{t_{0}}^{t+1}=\frac{nm_{r}}{t}\mathbf{B}_{r}+\frac{m_{s}}{t}(%
\mathbf{B}_{s}\odot\mathbf{B}_{r})\sum_{\lambda=t_{0}}^{t}\mathbf{P}%
_{t_{0}}^{\lambda}\ \ ,  \label{rtb4}
\end{equation}
where $\odot$ is the \emph{Hadamard product}: $\left(\mathbf{B}_{s}\odot%
\mathbf{B}_{r}\right)_{(ij)}\equiv\mathbf{B}_{s\ (ij)}\cdot\mathbf{B}_{r\
(ij)}$ .

From the decomposition given in equation (\ref{15}), it follows that
\begin{equation*}
\mathbf{B}_{s}\odot\mathbf{B}_{r}=\mathbf{B}_{s}\odot\left(\mathbf{QAQ}%
^{-1}\right)=\mathbf{Q}\left(\mathbf{B}_{s}\odot\mathbf{A}\right)\mathbf{Q}%
^{-1}\ \ ,
\end{equation*}
where the biases are such that $\mathbf{B}_{s}\odot\mathbf{A}$ is still a
Markov matrix.

The model with unbiased search, analyzed in previous sections, is a
special case of this model, where the matrix $\mathbf{B}_{s}$ is a matrix
of all $1$'s.

As we will see, weak integration still holds in the presence of biased
search. Moreover, while long-run integration generically does not hold,
partial integration occurs under an additional monotonicity condition. In
this case, while the distortions in type frequencies among an individual's
neighbors decrease over time, they do not vanish as the node accumulates
links.

\begin{proposition}
\label{prop-g1} If $m_{r}<1$, the general model with biased search satisfies
the \emph{weak integration property,} generically does not satisfy \emph{%
long-run integration}, and satisfies the \emph{partial integration property}
provided that $\mathbf{B}_{s}\odot\mathbf{A}$ satisfies monotone convergence.
\end{proposition}

\section{An empirical application to citation data}

\label{sec_citations}

In this section we use our random-search model to study the patterns of
cross-subfield scientific citations in physics.

The use of scientific citation data is motivated by several factors. First, there
is a literature that shows that key aspects of the time
evolution of citations can be captured by models based on variations of a
preferential attachment mechanism.  \cite{P76}, and then  \cite{R98} for ISI
papers and also \cite{N09}, found that older papers enjoy an
advantage in receiving citations, independently of the intrinsic
quality of the paper. Although a bias in favor of recent papers
allows for a better fit of certain datasets (see \cite{Berner_alii04,SR07}), the evidence of a
rich-gets-richer mechanism seems sound.  In addition, Refs. \cite{SR05,SR11} have shown that this evidence can be accounted for when
preferential attachment is generated by a random-search mechanism as the one
we use in this paper, where authors first ``randomly'' (from the econometrician's perspective)
select papers, and search these papers' reference lists to find additional citations.

There is less exploration of the patterns of citations across disciplines or
across other types of categories in which research may be organized. Several works have shown that geographical distance and national
boundaries are two important determinants of citation patterns, while \cite{LLJ04} has shown that citation patterns are quite
uniform across sub-fields in the high energy physics dataset (SPIRES). Also,
\cite{STA09} finds a relationship between the homophily in citing
other papers and the total citations received by computer science papers.

Thus, the generative process of citations is consistent with basic
aspects of the network formation process studied in this paper: First, it is
a growing network process, since papers appear in chronological
order, and old papers do not exit or die.  Second, citations are
directional, and only citations from newer to older nodes are possible.\footnote{There are revisions of papers
that allow them to cite relatively contemporaneous papers, but that seems to be a minor factor in the overall process.}
Third, citations cannot disappear, and accumulate over time. In addition, and
specifically to our model, nodes have ``types'', that we identify with the
scientific classification of a paper (see below for details). Finally, a key
element of our process is that links are formed both at random and by search
through established links. In the case of citations, these two channels of
search are present, since there is a difference between citations that come
from direct knowledge of a paper and citations that originate from the list
of references of other papers that one has read.

We use the American Physical Society (APS) citations dataset, which
reports papers published in journals of the APS between 1/1/1985 and 12/31/2003.\footnote{The journals in our data are Physical Review A, B, C, D , E, Letters, STPER and RevModPhys; considering
papers that have PACS classification codes which became compulsory in 1985 in the main six journals of the APS.
This dataset is available online from http://prola.aps.org/, and contains the data analyzed in \cite{RF11,R_alii09}.}
There is a total of 207912 papers and 1488866 citations (roughly 7 citations per paper
on average).
Around $42.8$ percent of the papers are never cited, while the
most cited one receives $952$ citations.

Types are defined by the first digit of the first (out of at most four) PACS classification code that characterize each paper:

\begin{itemize}
\item 00: General;

\item 10: The Physics of Elementary Particles and Fields;

\item 20: Nuclear Physics;

\item 30: Atomic and Molecular Physics;

\item 40: Electromagnetism, Optics, Acoustics, Heat Transfer, Classical
Mechanics, Fluid Dynamics;

\item 50: Physics of Gases, Plasmas, and Electric Discharges;

\item 60: Condensed Matter: Structural, Mechanical, and Thermal Properties;

\item 70: Condensed Matter: Electronic Structure, Electrical, Magnetic, and
Optical Properties;

\item 80: Interdisciplinary Physics and Related Areas of Science and
Technology;

\item 90: Geophysics, Astronomy, and Astrophysics.
\end{itemize}

\begin{figure}[htbp]
\begin{centering}
\includegraphics[width=12cm]{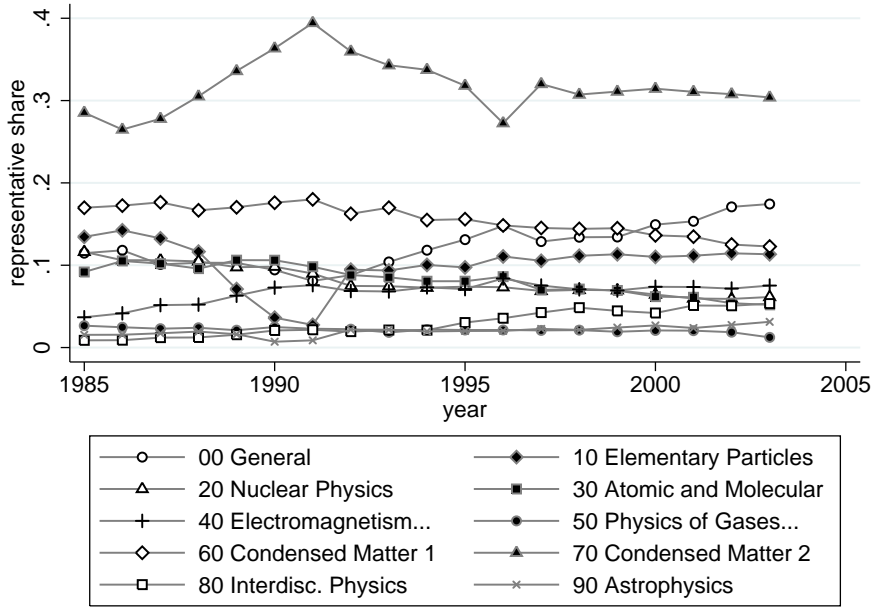}
\par\end{centering}
\caption{Shares of types' proportions over time}
\label{shares}
\end{figure}

We first note that the time profiles of types' population shares, measured,
for each type and for each year, as the proportion of the total papers
published during that year that are of that given type, is fairly
stationary during the whole period (see Figure \ref{shares}).\footnote{%
The only two sharp changes in the time profiles are around 1990 for type 10
(Physics of Elementary Particles and Fields) and type 70 (Condensed Matter
Electronic Structure, Electrical, Magnetic, and Optical Properties). These
changes were explained to us by the APS as follows (in private communication in response to our queries about these changes).
The increase of type 70 was driven by the sharp increase in the subcategory
74 ``Superconductivity'', due to the spike in interest in High Temperature Superconductivity that began in 1986 (including some switching of fields of high energy particle physicists some of whom
would have come from type 10).
The sharp decrease of type 10 was due to a 1989 policy by the APS
that increased page charges by 60 percent in the
Physical Review journals, including Physical Review D where much elementary particle/high energy physics is published.
Some authors reacted by publishing in other journals (outside of the APS data set) and increased use of the physics arXiv that started in 1991, and this reaction
was particularly heavy in the particle physics community. In 1996 APS removed page charges for properly prepared electronic manuscripts in Physical Review C and D, and in 1999 did the same for the other Physical Review journals. }
The approximate stationarity of most categories is roughly in line with our
assumption in the theoretical model that probabilities of birth of various
types are time invariant.  There are additional dramatic (unexplained) changes in the composition of fields following 2003, and so we truncate our data at that point.

In order to identify the various elements of our theoretical model, we need
to distinguish citations that originate from a direct random draw from the
pool of all existing papers (``random'' citations) from those that originate
from a search process that goes through the references contained in one's
random citations (``search'' citations). To do this, we proceed as follows.
We first identify a citation from paper A to paper C as a ``search''
citation if there exists some paper B with the following properties: 1) B is
published after C and before A, 2) A cites B, and 3) B cites C.

This method obviously has some degree of arbitrariness and will not
perfectly identify how the authors found the papers they cite. The bias of
this simplification is however not clear. On one side, it overstates the
weight of ``search'' in the citation process, since A may well cite C
because C is an important paper in the field, the reason for which also B also cites
C, without A having known about C though B. On the other side, however, it
could be that authors of paper A know about paper C only because they came
across paper B, which cites C: they could decide to cite only C because it
contains an older version of the same idea.  That is, it can be that some
papers are found through the search process, without the authors ever citing
the intermediate paper, and so some citations are coded as random even
though they were found through search. We stick with the strict
interpretation of the model, given that we have no other way of identifying
the actual process that the authors followed (see \cite{SR11} for an interesting strategy of identification).

Using this method we identify $56.34$ percent of total citations as ``search''
citations. We then classify the remaining $43.66$ percent of citations as
``random'' citations, being the complement of the ``search'' citations.

\subsection{Bias in random out-citations}

In order to identify the bias in the random part of the process, we compare
the share of ``random'' out-citations that are of the same type of the
citing paper with the population share of the type of the citing paper. The
first share ($q_{out}$ in table \ref{table_random}) is obtained by averaging
the share of random same-type out-citations of all papers of a given type
during the whole time period. The second share ($w$ in table \ref%
{table_random}) is obtained as the share of papers of a given type over all
papers in the sample for the whole time period.

\begin{table}[htp]
\begin{centering}
\begin{tabular}{r|rrrrrrrrrr}

Classification &         00 &         10 &         20 &         30 &         40 &         50 &         60 &         70 &         80 &         90 \\

\hline

Same type random cites: $q_{out}$ &       0.27 &       0.10 &       0.69 &       0.60 &       0.42 &       0.43 &       0.33 &       0.49 &       0.23 &       0.07 \\

Size of classification: $w$ &       0.13 &       0.10 &       0.08 &       0.08 &       0.07 &       0.02 &       0.15 &       0.32 &       0.03 &       0.02 \\

Coleman Index &       0.17 &       0.00 &       0.67 &       0.57 &       0.37 &       0.42 &       0.21 &       0.26 &       0.20 &       0.05 \\

\end{tabular}

\par\end{centering}
\caption{Same-type bias in the overall citations.}
\label{table_random}
\end{table}

The difference between these two shares is positive and substantial for all types, with
maximum value of about $.61$ for type $20$, minimum value of $.001$ for type $10$
(Interdisciplinary physics), and average value of $.26$.   Normalizing, for
each type, this difference by the the maximal potential difference given by
one minus the population share of the type, we obtain the Coleman (1958)
homophily index of each type ($ih=(q_{out}-w)/(1-w)$ in table \ref{table_random}).\footnote{%
This normalization has the purpose of allowing for meaningful comparison of
groups of different sizes, by taking into account the maximal potential
amount of homophily that each group has. See \cite{CJP09} for more discussion.} This index turns out not to be correlated with
types' population shares.

\subsection{Search bias, long--run integration, and partial integration}

One challenge with an empirical investigation of the various concepts of
integration is that certain papers happen to be intrinsically more cited
than others, simply because they are more fundamental or important than
others for their discipline. This type of ``fitness'' is independent of the
age of the paper, and is not modeled in our analysis.\footnote{%
See \cite{Atalay11} for the analysis of a single--type random--search process
which is based on fitness.} More importantly, it could potentially outweigh
the effect of time, and of the large in-degree that older nodes accumulate
in time, which is one of the forces behind the long-run integration property.

We deal with this problem by looking at the type-composition of the first $\tau$
citations of each paper, thereby replacing time with citation order.
This allows us to normalize the time--scale of each single paper, as if they
all had the same fitness. In this new context, which takes a form similar to that of Proposition \ref{Prop3}, the hypothesis we are testing
is whether the homophily of the in-degrees of a paper decreases with the
order of its in-citations. This is meant to capture the main
force that leads to partial integration: the growth of nodes' in-degree is
to a large extent composed of in-citations of the ``search'' type, which
in the case of unbiased search are less biased towards one's own type than
in-citations of the ``random'' kind.

A way to test partial integration is to measure the probability that a given citation originated from a paper in the same field as the cited paper; more precisely, we estimate the change in this probability associated with an increase in the order of the in-citation. The partial integration hypothesis predicts that this probability should decrease with the order of the in-citation.
We estimate a probit model where a dummy \emph{in--group} (taking value one if the citation comes from a paper in the same field as the cited paper and zero otherwise) is estimated as a function of the order of the citation.  The dataset contains 1034569 total citation, and we run separate regressions by looking at each sub-field separately (i.e., we look at all citations received by all papers belonging to each given sub-field). Results are reported in the next table.

\begin{table}[htdp]
\begin{center}\begin{tabular}{c|c|c}Type & DProb & $P>|z|$ \\\hline All Fields & -.00025** & 0.000 \\\hline 00 & .00028** & 0.000 \\\hline 10 & -.00067** & 0.000 \\\hline 20 & -00191** & 0.000 \\\hline 30 & -.00252**  & 0.000 \\\hline 40 & -.00034**  & 0.000 \\\hline 50 & .0008** & 0.009 \\\hline 60 & -.00112**  & 0.000 \\\hline 70 & -.00024** & 0.000 \\\hline 80 & -.00418**  & 0.000 \\\hline 90 & .00028  & 0.096 \end{tabular}
\end{center}
\caption{Expected change in probability of a citation being homophilous associated with an increase in the order of the citation. (**)=99\%.}
\label{probit}
\end{table}

We observe that for all sub-fields except type 00, type 50 and type 90, citations of higher order are less likely to come from papers within the same field. This is true also in the aggregate if we compute the expected change in probability without keeping track of the specific field of the receiving paper.  Types 00 and 50 behave differently and have an increase in the probability of homophilous citations, while type 90 has no significant change in probability.  At least broadly, this is consistent with partial integration.  On average, the probability of being cited by papers of the same field decreases by $0.025$\%: on average, the share of in-field citations decreases by 2.5 percentage points between the first and the 100th in-citation.

Beyond testing partial integration, we also examine whether the fraction of a paper's citations from random versus search becomes more tilted towards search as its number of citations grows, as would be consistent with the model.   This is plotted in Figure \ref{searchtime} below.

\begin{figure}[htbp]
\begin{centering}
\includegraphics[width=12cm]{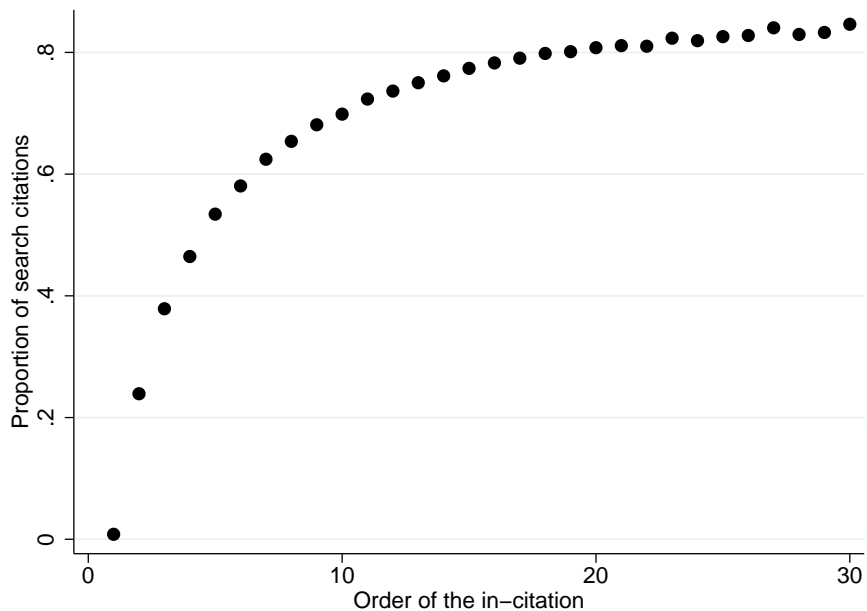}
\par\end{centering}
\caption{Average Fraction of Search Citations by Citation Order}
\label{searchtime}
\end{figure}

There is an evident upward trend, consistent with the model's predictions.

\section{Concluding remarks}

\label{sec_conclusion}

This paper contributes to our understanding of how heterogeneity and homophily among individuals impact the networks that they form. We have
built on the framework of Jackson and Rogers (2007), in which links result
both from meeting others at random and through introductions to their
neighbors, allowing both of these channels for link formation to be biased
by the types of the nodes involved. Some applications of interest have
significant type-based biases. Scholars are more likely to read papers from
their own field, people are more likely to befriend those with a similar
background, organizations have closer ties within departments, and so on. We
do not attempt here to model the source of these biases, but take the model
as a reduced form representation of the resulting effects that such biases have on how links are
formed.

Within the context of this framework, long-run integration, whereby old
nodes obtain local networks that asymptotically resemble the population at
large, occurs if and only if the search part of the network formation
process (the second channel mentioned above) is type-unbiased so that the
only bias in the process comes in which nodes are initially found through the
random meeting process.

Understanding the neighborhoods of old nodes is important since these nodes
constitute the hubs of the network. If one is interested in processes
occurring on the network such as, e.g., strategic behavior or diffusion
processes, then the characteristics of hub nodes are of central importance.
On the other hand, there are many other important properties of the network
that may be affected by type-based biases. In order to analyze these
properties, we turned to the more specific model of location-based biases
among two groups, deriving implications regarding type-based degree
distributions and on group-level homophily.

We leave open a number of interesting questions. First, there is the matter
of exploring the extent to which the results from the location-based biases
generalize. Second, there are other kinds of network formation processes in
which similar questions could be addressed. In fact, in the different
model of \cite{CJP09,CJP10a}, there are some similar
results, but in general we have an incomplete understanding of how
heterogeneity impacts network formation. Third, there are many summary
statistics of networks that can be generalized to a multi-type framework,
including clustering measures, that can be analyzed in future work.

\bigskip{}

\setcounter{section}{1}

\appendix
\global\long\def\thesection{Appendix \Alph{section}}
 \global\long\def\thesubsection{\Alph{section}.\arabic{subsection}}
 \setcounter{equation}{0} \global\long\def\theequation{\alph{equation}}
 \setcounter{theorem}{0} \global\long\def\thetheorem{\Alph{theorem}}

\section{Some results on Markov matrices}

\label{matrices}

This first Section of the Appendix provides some results that are necessary
for the proofs of our results. Take an $H\times H$ Markov matrix $\mathbf{M}$
with all positive elements, i.e. a positive Markov matrix.

\begin{lemma}
\label{matrix_markov} For every $x>0$ the $H\times H$ matrix
\begin{equation*}
\mathbf{M}(x)\equiv\left(e^{x}-1\right)^{-1}\sum_{\mu=1}^{\infty}\frac{%
x^{\mu}}{\mu!}\mathbf{M}^{\mu}=\frac{\mbox{exp}~(\mathbf{M}x)-\mathbf{I}}{%
\mbox{exp}~(x)-1}
\end{equation*}
is a Markov matrix.
\end{lemma}

\textbf{Proof:} for every $\mu\in\mathbb{N}$, $\mathbf{M}^{\mu}$ is a Markov
matrix. To show that $\mathbf{M}(x)$ is a Markov matrix, we need to prove
that for every $i,j\in\{1,\dots,H\}$ we have that $0<M(x)_{ij}<1$, and that $%
\sum_{k=1}^{H}M(x)_{kj}=1$. \newline
The first condition comes from the fact that $M(x)_{ij}$ is a convex
combination of (an infinite number of) probabilities. \newline
The second condition comes from the fact that
\begin{equation*}
\sum_{k=1}^{H}[M(x)]_{kj}=\left(e^{x}-1\right)^{-1}\sum_{\mu=1}^{\infty}%
\frac{x^{\mu}}{\mu!}\left(\sum_{k=1}^{H}\left[M^{\mu}\right]%
_{kj}\right)=\left(e^{x}-1\right)^{-1}\sum_{\mu=1}^{\infty}\frac{x^{\mu}}{%
\mu!}=1\ \ .\mbox{\eproof}
\end{equation*}

\noindent \bigskip{}

$\mathbf{M}(x)$ can be seen as a weighted average of the infinite elements
of $\{\mathbf{M}^{\mu}\}_{\mu\in\mathbb{N}}$. \newline
We know that
\begin{eqnarray}
\lim_{\mu\rightarrow\infty}\mathbf{M}^{\mu} & = & \left(%
\begin{array}{c}
\vec{v}(\mathbf{M})^{\prime } \\
\vdots \\
\vec{v}(\mathbf{M})^{\prime }%
\end{array}%
\right)\ \ ,
\end{eqnarray}
where the row--vector $\vec{v}(\mathbf{M})^{\prime }$ is the unique
eigenvector associated with eigenvalue $1$ of matrix $\mathbf{M}$ (up to a
normalization that it's elements sum to one, by the Perron--Frobenius
Theorem). We define this matrix at the limit, with all equal elements on
each column, as $\bar{\mathbf{M}}$. Now we prove a relation between the
limit of $\mathbf{M}(x)$ and $\bar{\mathbf{M}}$.

\begin{lemma}
\label{matrix_limit} For every positive Markov matrix $\mathbf{A}$, and for
every couple $i,j\in\{1,\dots,H\}$, we have that
\begin{equation*}
\lim_{x\rightarrow\infty}[M(x)]_{ij}=\lim_{\mu\rightarrow\infty}[M^{%
\mu}]_{ij}=[\bar{M}]_{ij}\ \ .
\end{equation*}
\end{lemma}

\noindent \textbf{Proof:} By definition of $\bar{\mathbf{M}}$, for every $%
\epsilon>0$ there is a number $\bar{k}\in\mathbb{N}$, such that for every $%
\mu>\bar{k}$, we have $\Big|\lbrack M^{\mu}]_{ij}-[M]_{ij}\Big|<\epsilon$.
By driving $x\rightarrow\infty$ we can impose to $0$ the weight $\frac{%
x^{\nu}}{(e^{x}-1)\nu!}$ of every $\nu<\bar{k}$. In this way $[M(x)]_{ij}$
becomes a weighted average of almost only elements from $\{M^{\mu}\}_{\mu\in%
\mathbb{N}}$, with $\mu>\bar{k}$. As for all of them we have $\Big|\lbrack
M^{\mu}]_{ij}-[\bar{M}]_{ij}\Big|<\epsilon$, we have the result. %
\hbox{\hskip3pt\vrule width4pt height8pt depth1.5pt}

\bigskip{}

\begin{definition}
$\mathbf{M}$ satisfies the \emph{monotone convergence property} if, for
every couple $i,j\in\{1,\dots,H\}$, and for every $\mu\in\mathbb{N}$, the
element $M_{ij}^{\mu}$ has the following properties:

\begin{enumerate}
\item if $M_{ij}>\bar{M}_{ij}$, then $M_{ij}\geq M_{ij}^{\mu}\geq
M_{ij}^{\mu+1}\geq\bar{M}_{ij}$;

\item if $M_{ij}<\bar{M}_{ij}$, then $M_{ij}\leq M_{ij}^{\mu}\leq
M_{ij}^{\mu+1}\leq\bar{M}_{ij}$.
\end{enumerate}
\end{definition}

What comes out directly from the definition is that, if $M_{ij}>\bar{M}_{ij}$%
, then there is at least one $\mu$ for which the inequality is strict, i.e. $%
M_{ij}^{\mu}>M_{ij}^{\mu+1}$.

\begin{lemma}
\label{matrix_monotone} For every couple $i,j\in\{1,\dots,H\}$, and for
every $x>0$ If $\mathbf{M}$ satisfies the \emph{monotone convergence property%
}, then

\begin{enumerate}
\item if $M_{ij}>\bar{M}_{ij}$, then $\frac{\partial}{\partial x}%
[M(x)]_{ij}<0$;

\item if $M_{ij}<\bar{M}_{ij}$, then $\frac{\partial}{\partial x}%
[M(x)]_{ij}>0$.
\end{enumerate}
\end{lemma}

\noindent \textbf{Proof:} We focus on case 1, as the other is proven by
reversing inequalities.

First, note that the function
\begin{equation*}
\frac{\mu}{x}\left(e^{x}-1\right)-e^{x}
\end{equation*}
is negative if and only if
\begin{equation*}
\mu<\frac{xe^{x}}{e^{x}-1}\ \ .
\end{equation*}
Let us call $\nu(x)$ the minimum integer strictly above $\frac{xe^{x}}{%
e^{x}-1}$, i.e. $\nu(x)\equiv\lceil\frac{xe^{x}}{e^{x}-1}\rceil$.

Now we can show that
\begin{eqnarray}
\frac{\partial}{\partial x}[M(x)]_{ij} & = & \frac{1}{\left(e^{x}-1%
\right)^{2}}\sum_{\mu=1}^{\infty}\frac{x^{\mu}}{\mu!}\left(\frac{\mu}{x}%
\left(e^{x}-1\right)-e^{x}\right)\left[M^{\mu}\right]_{ij}  \notag \\
& < & \frac{1}{\left(e^{x}-1\right)^{2}}\sum_{\mu=1}^{\infty}\frac{x^{\mu}}{%
\mu!}\left(\frac{\mu}{x}\left(e^{x}-1\right)-e^{x}\right)\left[M^{\nu(x)}%
\right]_{ij}  \label{derivative_monotone}
\end{eqnarray}
It is a matter of calculus to check that
\begin{equation*}
\sum_{\mu=1}^{\infty}\frac{x^{\mu}}{\mu!}\left(\frac{\mu}{x}%
\left(e^{x}-1\right)-e^{x}\right)=0\ \ ,
\end{equation*}
and then the derivative in (\ref{derivative_monotone}) is strictly negative. %
\hbox{\hskip3pt\vrule width4pt height8pt depth1.5pt}

\bigskip{}

Finally, we provide a simple sufficient condition for a $2\times2$ Markov
matrix.

\begin{lemma}
\label{matrix_2x2} Consider the $2\times2$ Markov matrix $\left(%
\begin{array}{cc}
p & 1-p \\
1-q & q%
\end{array}%
\right)$, if $p+q\geq1$, then it has the \emph{monotone convergence property}%
.
\end{lemma}

\noindent \textbf{Proof:} By the Perron--Frobenius Theorem this matrix
converges to $\left(%
\begin{array}{cc}
u & 1-u \\
u & 1-u%
\end{array}%
\right)$, with $u=\frac{1-q}{(1-p)+(1-q)}$.

It is easy to check that, as $p\geq1-q$, then $p\geq u$. By symmetry between
$p$ and $q$, also $q\geq1-u$. \newline
Now consider a matrix $\left(%
\begin{array}{cc}
\alpha & 1-\alpha \\
1-\beta & \beta%
\end{array}%
\right)$, such that $p\geq\alpha\geq u$ and $q\geq\beta\geq1-u$. To finish
the proof it is enough to show that
\begin{equation}
p\geq\alpha p+(1-\alpha)(1-q)\geq u\ \ ,  \label{ineq_2x2}
\end{equation}
as it will be proved symmetrically also with respect to $q$ and $\beta$. The
middle term of these inequalities is increasing in $\alpha$, as $p\geq1-q$.
If $\alpha=u$ it is equal to $u$, if instead $\alpha=p$, with some algebraic
substitution, we have that again both inequalities are satisfied, as $%
p\geq1-q$. This completes the proof.
\hbox{\hskip3pt\vrule width4pt
height8pt depth1.5pt}

\section{Proofs}

\label{proofs}

\subsection{Proofs for Section \protect\ref{sec_no_bias_search}.}

\noindent \textbf{Proof of Proposition \ref{prop_1} (page \pageref{prop_1}):}
Note first that the node born at time $t^{\prime}$ in definition \ref{weak}
has, at the beginning of time $t^{\prime}+1$ (before node $t^{\prime}+1$
sends its links) an in-degree of $0$. This directly implies that the
probability of $t^{\prime}$ to receive a link at time $t^{\prime}+1$ from a
node of type $\theta^{\prime \prime }$, given that such a node is born, is
equal to the probability of being found at random among the $t^{\prime}$
nodes in the network. This probability is equal to:
\begin{equation}
\frac{nm_{r}}{t^{\prime}}p(\theta^{\prime \prime \prime}))\ .  \label{PI1}
\end{equation}
On the other hand, the probability that node $t_{0}$ is be found at time $%
t^{\prime}+1$ is the sum of the probability of being found at random and
through search. In the model with homogeneous search, this is:
\begin{equation}
\frac{nm_{r}}{t^{\prime}}p(\theta^{\prime \prime
},\theta(t_{0}))+nm_{s}\sum_{\theta\in\Theta}p(\theta^{\prime \prime
},\theta)\frac{\Pi_{t_{0}}^{t^{\prime}}(\theta,\theta(t_{0}))}{%
t^{\prime}p(\theta(t_{0}))}\frac{1}{n}\ .  \label{PI2}
\end{equation}
Note that in (\ref{PI2}) the terms in the vector $\Pi_{t_{0}}^{t^{\prime}}(%
\theta,\theta(t_{0}))$ grow without bound as $t^{\prime}$ tends to infinity,
while the first terms in (\ref{PI2}) and in (\ref{PI1}) are constant once $%
t^{\prime}$ is eliminated from the denominator of both expressions. It
follows that we can always choose a $t^{\prime}$ large enough for (\ref{PI2}%
) to be larger than (\ref{PI1}).
\hbox{\hskip3pt\vrule width4pt
height8pt depth1.5pt}

\bigskip{}

\noindent \textbf{Proof of Proposition \ref{prop_lri} (page \pageref%
{prop_lri}):} We want to see how the matrix $\Pi_{t_{0}}^{t}$ of
type--by--type links for a node born at time $t_{0}$ develops. To do this we
compare its behavior with the behavior of the type--blind process, where the
in--links evolve according to\footnote{%
This process reduces to the $1$--type case studied in Jackson and Rogers
(2007).}
\begin{equation*}
\pi_{t_{0}}(t)=n\frac{m_{r}}{m_{s}}\left(\left(\frac{t}{t_{0}}%
\right)^{m_{s}}-1\right)\ \ .
\end{equation*}
To make this comparison in the long run we study
\begin{equation}
\lim_{t\rightarrow\infty}\frac{\Pi_{t_{0}}^{t}}{\pi_{t_{0}}(t)}\ \ .
\label{limit}
\end{equation}

\bigskip{}

Consider the solution of the model, as described by equation (\ref{13}),
with the decomposition $\mathbf{B}_{r}=\mathbf{QAQ}^{-1}$.

We rewrite (\ref{13}) as:
\begin{equation*}
\Pi_{t_{0}}^{t}=\frac{m_{r}}{m_{s}}\left(\left(\frac{t}{t_{0}}\right)^{m_{s}%
\mathbf{QAQ}^{-1}}-\mathbf{I}\right).
\end{equation*}
We now use some results from Section \ref{sec_model_dynamic}. By equation (%
\ref{15}), and the facts that $\mathbf{I=QIQ}^{-1}$ and $\mathbf{A}^{n}=%
\mathbf{QA^{n}Q}^{-1}$, we obtain:
\begin{eqnarray}
\Pi_{t_{0}}^{t} & = & n\frac{m_{r}}{m_{s}}\mathbf{Q}\left(\left(\sum_{%
\mu=0}^{\infty}\frac{\left(m_{s}\log\left(\frac{t}{t_{0}}\right)\mathbf{A}%
\right)^{\mu}}{\mu!}\right)-\mathbf{I}\right)\mathbf{Q}^{-1}  \notag \\
& = & n\frac{m_{r}}{m_{s}}\left(\left(\frac{t}{t_{0}}\right)^{m_{s}}-1\right)%
\mathbf{Q}\left(\left(\left(\frac{t}{t_{0}}\right)^{m_{s}}-1\right)^{-1}%
\sum_{\mu=1}^{\infty}\frac{\left(m_{s}\log\left(\frac{t}{t_{0}}%
\right)\right)^{\mu}}{\mu!}\mathbf{A}^{\mu}\right)\mathbf{Q}^{-1}\ \ .
\label{19}
\end{eqnarray}

Limit (\ref{limit}) implies that (we use Lemma \ref{matrix_limit} from \ref%
{matrices})
\begin{eqnarray}
\lim_{t\rightarrow\infty}\frac{\Pi_{t_{0}}^{t}}{\pi(t)} & = &
\lim_{t\rightarrow\infty}\left(\mathbf{Q}\left(\left(\left(\frac{t}{t_{0}}%
\right)^{m_{s}}-1\right)^{-1}\sum_{\mu=1}^{\infty}\frac{\left(m_{s}\log\left(%
\frac{t}{t_{0}}\right)\right)^{\mu}}{\mu!}\mathbf{A}^{\mu}\right)\mathbf{Q}%
^{-1}\right)  \notag \\
& = & \mathbf{Q}\left(\lim_{\mu\rightarrow\infty}\mathbf{A}^{\mu}\right)%
\mathbf{Q}^{-1}  \notag \\
& = & \mathbf{Q}\left(%
\begin{array}{c}
\vec{v}(\mathbf{A})^{\prime } \\
\vdots \\
\vec{v}(\mathbf{A})^{\prime }%
\end{array}%
\right)\mathbf{Q}^{-1}\ \ ,
\end{eqnarray}
where the row--vector $\vec{v}(\mathbf{A})^{\prime }$ is the unique
eigenvector associated with eigenvalue $1$ of matrix $\mathbf{A}$
(normalized to sum to 1). In this way, in the long run a node of type $i$
born at time $t_{0}$ receives a fraction of in--links from nodes of type $j$
which is given by the ratio
\begin{equation*}
p(j)\frac{\vec{v}(\mathbf{A})_{i}}{p(i)}
\end{equation*}
of the overall nodes that it would receive in a type--blind process. This
proportion is the product of $p(j)$ times a term that is constant for type $%
i $. \hbox{\hskip3pt\vrule width4pt height8pt depth1.5pt}

\bigskip{}

\noindent \textbf{Proof of Proposition \ref{prop_partial_no_bias} (page %
\pageref{prop_partial_no_bias}):} The result comes from the expression of
matrix $\Pi_{t_{0}}^{t}$ as defined in equation (\ref{19}), in the Proof of
Proposition \ref{prop_lri}:
\begin{equation*}
\frac{\Pi_{t_{0}}^{t}}{\pi_{t_{0}}(t)}=\mathbf{Q}\left(\left(\left(\frac{t}{%
t_{0}}\right)^{m_{s}}-1\right)^{-1}\sum_{\mu=1}^{\infty}\frac{%
\left(m_{s}\log\left(\frac{t}{t_{0}}\right)\right)^{\mu}}{\mu!}\mathbf{A}%
^{\mu}\right)\mathbf{Q}^{-1}\ \ .
\end{equation*}
Here $\left(\left(\frac{t}{t_{0}}\right)^{m_{s}}-1\right)^{-1}$ is just a
rescaling term so that the matrix in brackets is again a Markov matrix (see
Lemma \ref{matrix_markov} in \ref{matrices}).
From the Proof of Proposition \ref{prop_lri} we know that it converges to
the distribution of the population shares. As $\mathbf{A}$ satisfies the
monotone convergence property, we can apply Lemma \ref{matrix_monotone} from %
\ref{matrices} to prove that this convergence is monotonic. %
\hbox{\hskip3pt\vrule width4pt height8pt depth1.5pt}

\bigskip{}

\noindent \textbf{Proof of Proposition \ref{prop_outdegree} (page \pageref%
{prop_outdegree}):}

Expressing the steady state equation (\ref{steady_state_outdegree}) we
obtain
\begin{equation}
\mathbf{\bar{D}}=(1-m_{s})\left(\mathbf{I}-m_{s}\mathbf{B}_{r}\right)^{-1}%
\mathbf{B}_{r}\ \ .
\end{equation}

Using the algebraic identity
\begin{equation*}
\left(\mathbf{I}-m_{s}\mathbf{B}_{r}\right)^{-1}=\sum_{\mu=0}^{\infty}%
\left(m_{s}\mathbf{B}_{r}\right)^{\mu}\ \ ,
\end{equation*}
we obtain the following expression: 
\begin{equation*}
{\bar{D}}=\mathbf{B}\left(\frac{1-m_{s}}{m_{s}}\sum_{\mu=1}^{\infty}%
\left(m_{s}\mathbf{A}\right)^{\mu}\right)\mathbf{B}^{-1}
\end{equation*}
.

In the above expression, the matrix in brackets is such that, as $%
m_{s}\rightarrow1$, the elements of each column homogenize (see Lemma \ref%
{matrix_limit} of \ref{matrices}). However, full homogeneity only occurs at
the limit $m_{s}\rightarrow1$.

To obtain some insight on the time evolution of the out-degree, let us
express equation (\ref{out_degree_diff_eq}) as a differential equation, and
solve it explicitly (as we have done in (\ref{13}) for the in-degree).

The system is
\begin{eqnarray}
\frac{\partial}{\partial t}\mathbf{\Delta}_{t} & = & (1-m_{s})\mathbf{B}%
_{r}+m_{s}\mathbf{B}_{r}\frac{\mathbf{\Delta}_{t}}{t}\ \ ,
\end{eqnarray}
with solution:
\begin{equation*}
\mathbf{\Delta}_{t}=\bar{\mathbf{D}}t+\mathbf{C}t^{m_{s}\mathbf{B}_{r}}\ \ ,
\end{equation*}

where $\mathbf{C}$ is a constant matrix.

For a given initial condition $\mathbf{D}_{1}$ (that we can identify with
the matrix $\mathbf{A}$ of biases) the solution for $\mathbf{D}_{t}$ can be
written as:
\begin{eqnarray}
\mathbf{D}_{t} & = & \frac{\partial}{\partial t}\mathbf{\Delta}_{t}=\bar{%
\mathbf{D}}+\frac{1}{t}(\mathbf{D}_{1}-\bar{\mathbf{D}})t^{m_{s}\mathbf{B}%
_{r}}\ \ .
\end{eqnarray}
\hbox{\hskip3pt\vrule width4pt height8pt depth1.5pt}

\subsection{Proofs for Section \protect\ref{sec_location_bias}.}

For the proofs of this section we need an additional preliminary result,
that follows here below

Consider a degree distribution $F(k)$ obtained implicitly through a process
such that the growth of a node born at $t_{0}$ is governed by $%
k_{t_{0}}(t)=f(t/t_{0})$ and another degree distribution $G$ such that $%
k_{t_{0}}(t)=g(t/t_{0})$, with $k_{t_{0}}(t_{0})=0$. Assume $f$ and $g$ are
weakly increasing and continuous on $[1,+\infty\lbrack$ and that $%
\lim_{x\rightarrow\infty}f(x)=\lim_{x\rightarrow\infty}g(x)=\infty$.

\begin{lemma}
\label{fosd} $F$ First-Order Stochastically Dominates $G$ if and only if for
all $x\geq1,f(x)\geq g(x)$, with strict inequality for some $x$.
\end{lemma}

\noindent \textbf{Proof of Lemma \ref{fosd}:} Assume $f(x)\geq g(x)$ for all
$x\geq1$. Pick $k$ and $t$ arbitrarily. Define $i_{f}$ as the birthdate of
the node with degree $k$ at time $t$ under $f$, and similarly for $i_{g}$.
We have $f(t/i_{f})=k\geq g(t/i_{f})$, which, since $g$ is non-decreasing
implies that $i_{g}\leq i_{f}$. Since $F_{t}(k)=1-i_{f}/t$ and $%
G_{t}(k)=1-i_{g}/t$, we have $F_{t}(k)\leq G_{t}(k)$.

Now take $\bar{x}$ such that $f(\bar{x})>g(\bar{x})$. Pick $k$ and $t$
arbitrarily. Define $i_{f}=t/\bar{x}$ and $\bar{k}$ to be the size of node $%
i_{f}$ at time $t$ under $f$. Then set $i_{g}$ to be the node with degree $%
\bar{k}$ at time $t$ under $g$. We have $\bar{k}=f(t/i_{f})=f(\bar{x})>g(%
\bar{x})=g(t/i_{f})$, which implies that $i_{g}<i_{f}$. Thus $F_{t}(\bar{k}%
)<G_{t}(\bar{k})$.

To show necessity, fix $t$ and choose $i_{f}$ so that $f(t/i_{f})<g(t/i_{f})$%
, and set $\bar{k}=f(t/i_{f})$. Defining $i_{g}$ as the node with degree $%
\bar{k}$ at time $t$ under $f$, we know that $i_{g}>i_{f}$. This implies
that $G_{t}(\bar{k})<F_{t}(\bar{k})$, completing the proof. %
\hbox{\hskip3pt\vrule width4pt height8pt depth1.5pt}

\bigskip{}

Now we can proceed with the proofs of Section \ref{sec_location_bias}.

\bigskip{}

\noindent \textbf{Proof of Lemma \ref{L1} (page \pageref{L1}):} Apply to
this particular case the expression from (\ref{13}), considering the
decomposition from (\ref{15}) :
\begin{equation*}
\Pi_{t_{0}}^{t}=n\frac{m_{r}}{m_{s}}\left(\mathbf{Q}\left(\frac{t}{t_{0}}%
\right)^{m_{s}\left(%
\begin{array}{cc}
p_{1} & 1-p_{1} \\
1-p_{2} & p_{2}%
\end{array}%
\right)}\mathbf{Q}^{-1}-\mathbf{I}\right)\ \ ,
\end{equation*}
where we have called $p_{1}\equiv p(\theta_{1},\theta_{1})$ and $p_{2}\equiv
p(\theta_{2},\theta_{2})$.

As can be directly computed checking \cite{MJZ07}, we have that
\begin{equation*}
\left(\frac{t}{t_{0}}\right)^{m_{s}\left(%
\begin{array}{cc}
p_{1} & 1-p_{1} \\
1-p_{2} & p_{2}%
\end{array}%
\right)}=\left(%
\begin{array}{cc}
\frac{(1-p_{1})t^{m_{s}(p_{1}+p_{2}-1)}+(1-p_{2})t^{m}}{2-p_{1}-p_{2}} &
\dots \\
&  \\
\frac{(1-p_{2})\left(t_{s}^{m}-t^{m_{s}(p_{1}+p_{2}-1)}\right)}{2-p_{1}-p_{2}%
} & \dots%
\end{array}%
\right)\ \ ,
\end{equation*}
where we have reported only the first column.

Recall that $p_{1}=\frac{p\gamma^{2}}{(1-p)(1-\gamma)+p\gamma}+\frac{%
p(1-\gamma)^{2}}{p(1-\gamma)+(1-p)\gamma}$ and $p_{2}=\frac{(1-p)\gamma^{2}}{%
p(1-\gamma)+(1-p)\gamma}+\frac{(1-p)(1-\gamma)^{2}}{(1-p)(1-\gamma)+p\gamma}$%
. Some manipulations show that $\frac{1-p_{1}}{2-p_{1}-p_{2}}=1-p$, and
symmetrically $\frac{1-p_{2}}{2-p_{1}-p_{2}}=p$.\footnote{$\left(\frac{%
1-p_{1}}{2-p_{1}-p_{2}},\frac{1-p_{2}}{2-p_{1}-p_{2}}\right)$ is actually
the eigenvector of $\left(%
\begin{array}{cc}
p_{1} & 1-p_{1} \\
1-p_{2} & p_{2}%
\end{array}%
\right)$ associated to its asymptotic limit $\lim_{t\rightarrow\infty}\left(%
\begin{array}{cc}
p_{1} & 1-p_{1} \\
1-p_{2} & p_{2}%
\end{array}%
\right)^{t}$ (see the proof of Lemma \ref{matrix_2x2} in \ref{matrices}).
Considering the location--based model, it is reasonable that this limit does
not depend on $\gamma$ but only on the initial distribution of the two
types, given by $p$.} If we finally consider that $\mathbf{Q}=\left(%
\begin{array}{cc}
p & 0 \\
0 & 1-p%
\end{array}%
\right)$, we have the result.
\hbox{\hskip3pt\vrule width4pt height8pt
depth1.5pt}

\bigskip{}

\noindent \textbf{Proof of Proposition \ref{Prop3} (page \pageref{Prop3}):}
We take the case of $\theta_{1}$. The case of $\theta_{2}$ is analogous.
Define $f(x)=nm_{r}/m_{s}(x^{m_{s}}-1)$; hence $f^{-1}(k)=(1+\frac{km_{s}}{%
nm_{r}})^{1/m_{s}}$. Next define $g(x)=p+(1-p)(x^{bm_{s}}-1)/(x^{m_{s}}-1)$.
Notice that at time $t$, the proportion of in-links that a node of type $%
\theta_{1}$ born at time $t_{0}<t$ has from its own group is
\begin{equation*}
\frac{\Pi_{t_{0}}^{t}(1,1)}{\Pi_{t_{0}}^{t}(1,1)+\Pi_{t_{0}}^{t}(2,1)}.
\end{equation*}
It then follows from Lemma 1 that $r^{1}(k)=g(f^{-1}(k))$. Evaluating this
formula delivers the claimed expression.

Without loss of generality, we can set $nm_{r}/m_{s}=1$ in what follows.
Introduce $y=1+k$. Next, $r^{\prime\prime}=[(f^{-1})^{\prime}]^{2}g^{\prime%
\prime}\circ f^{-1}+(f^{-1})^{\prime\prime}g^{\prime}\circ f^{-1}$.
Developing and substituting shows that $r^{\prime\prime}$ has the same sign
as
\begin{equation*}
\varphi(y)=y^{b+2}(1-b)(2-b)+y^{b+1}2b(2-b)-y^{b}b(1-b)-2y^{2}
\end{equation*}
A detailed study of $\varphi$ and its first three derivatives then shows
that $\varphi(y)>0$ if $y>1$, hence that $r^{\prime\prime}(k)>0$ if $k>0$.

The explicit expressions for the derivative with respect to $p$ are not
trivial given that $b$ is a function of $p$. We have
\begin{eqnarray*}
\frac{\partial r}{\partial k} & = & -(1-p)\frac{(1+k)^{b-1}(1+(1-b)k)-1}{%
k^{2}} \\
k^{2}(1+k)^{1-b}\frac{\partial^{2}r}{\partial k\partial p} & = &
\psi(k)=1+(1-b)k-(1+k)^{1-b}+(1-p)(-\frac{\partial b}{\partial p})\left[%
\ln(1+k)(1+(1-b)k)-k\right]
\end{eqnarray*}
Also, note that $r(0)=p+(1-p)b$. Thus, $\frac{\partial r}{\partial p}(0)=1-b+%
\frac{\partial b}{\partial p}(1-p)$. We have: $b=1-\frac{\gamma(1-\gamma)}{%
p(1-p)(2\gamma-1)^{2}+\gamma(1-\gamma)}$ and $\frac{\partial b}{\partial p}=-%
\frac{(2p-1)(2\gamma-1)^{2}\gamma(1-\gamma)}{\left[p(1-p)(2\gamma-1)^{2}+%
\gamma(1-\gamma)\right]^{2}}$. Developing, we get that $\frac{\partial r}{%
\partial p}(0)$ has the same sign as $1-(2\gamma-1)^{2}p(2-p)-3\gamma(1-%
\gamma)\geq1-(2\gamma-1)^{2}-3\gamma(1-\gamma)=\gamma(1-\gamma)>0$ where the
first inequality comes from the fact that $p(2-p)\leq1$. Thus, $\frac{%
\partial r}{\partial p}(0)>0$ and $1-b>(-\frac{\partial b}{\partial p})(1-p)$%
.

Next, derive the function $\psi$ with respect to $k$. We have:

$\psi^{\prime}(k)=(1-b)(1-(1+k)^{-b})+(1-p)(-\frac{\partial b}{\partial p})%
\left[(1-b)\ln(1+k)-b(1-\frac{1}{1+k})\right]$ and

$(1+k)\psi^{\prime\prime}(k)=(1-p)(-\frac{\partial b}{\partial p}%
)(1-b)-(1-p)(-\frac{\partial b}{\partial p})b(1+k)^{-1}+b(1-b)(1+k)^{-b}$

Here, $\psi^{\prime\prime}(0)=b(1-b)+(1-p)(-\frac{\partial b}{\partial p}%
)(1-2b)$. Since $1-b>(-\frac{\partial b}{\partial p})(1-p)$, we have $%
\psi^{\prime\prime}(0)\geq(1-p)(-\frac{\partial b}{\partial p})(1-b)>0$.
Also, $\lim_{k\rightarrow\infty}\psi^{\prime\prime}(k)=0^{+}$. By looking at
its derivative, we see that the function $(1+k)\psi^{\prime\prime}(k)$ is
either decreasing, or increasing and decreasing. In either case, since it is
positive when $k=0$ and when $k\rightarrow\infty$, it must be greater than
or equal to zero for any $k$. Thus, $\psi^{\prime\prime}>0$ if $k>0$ hence $%
\psi^{\prime}$ is increasing. Since $\psi^{\prime}(0)=0$, $\psi^{\prime}>0$
if $k>0$. Thus, $\psi$ is increasing and as $\psi(0)=0$, $\psi>0$ and $\frac{%
\partial^{2}r}{\partial k\partial p}>0$ if $k>0$ . Finally, given that $%
\frac{\partial r}{\partial p}(0)>0$ and $\frac{\partial r}{\partial p}$ is
increasing in $k$, $\frac{\partial r}{\partial p}>0$, $\forall k$. %
\hbox{\hskip3pt\vrule width4pt height8pt depth1.5pt}

\bigskip{}

\noindent \textbf{Proof of Proposition \ref{3part prop} (page \pageref{3part
prop}):} For (i), use Lemma \ref{L1} to write
\begin{eqnarray*}
\Pi_{t_{0}}^{t}(1,1) & = & n\frac{m_{r}}{m_{s}}p\left[(\frac{t}{t_{0}}%
)^{m_{s}}-(\frac{t}{t_{0}})^{bm_{s}}\right]+n\frac{m_{r}}{m_{s}}\left[(\frac{%
t}{t_{0}})^{bm_{s}}-1\right] \\
\Pi_{t_{0}}^{t}(2,1) & = & n\frac{m_{r}}{m_{s}}(1-p)\left[(\frac{t}{t_{0}}%
)^{m_{s}}-(\frac{t}{t_{0}})^{bm_{s}}\right].
\end{eqnarray*}
Given that $p>1/2$ and that $b\geq0$, we know that $\frac{1-p}{p}<1$ and the
second term in the first equation is non-negative. Thus $%
\Pi_{t_{0}}^{t}(1,1)>\Pi_{t_{0}}^{t}(2,1)$ for all $t\geq t_{0}$, which
allows us to apply Lemma \ref{fosd}.

Now consider the expressions for $\Pi_{t_{0}}^{t}(2,2)$ and $%
\Pi_{t_{0}}^{t}(1,2)$ obtained from the above equations by exchanging $p$
with $1-p$. When $p>1/2$ (meaning $\theta_{1}$ is the majority group) then $%
\frac{p}{1-p}>1$, and when $b<1$ (i.e., $\gamma<1$, meaning there is at
least some inter-group linking) then for large values of $t/t_{0}$ the
second term in the expression for $\Pi_{t_{0}}^{t}(2,2)$ becomes negligible,
in which case $\Pi_{t_{0}}^{t}(2,2)<\Pi_{t_{0}}^{t}(1,2)$ in the upper tail,
proving (ii) by application of Lemma \ref{fosd}.

For (iii), introduce the function $%
\psi(x)=p(1-p)[x^{m_{s}}-x^{bm_{s}}]-(1-p)[(1-p)x^{m_{s}}+px^{bm_{s}}-1]$.
Note that $\psi(t/t_{0})\geq0$ if and only if $\Pi_{t_{0}}^{t}(1,2)\geq%
\Pi_{t_{0}}^{t}(2,2)$. Observe that $\psi(1)=0$. Also,

\begin{equation*}
\psi^{\prime}(x)=x^{m_{s}-1}\left[%
(2p-1)(1-p)m_{s}-2p(1-p)bm_{s}x^{(b-1)m_{s}}\right]
\end{equation*}
Since $b-1\leq0$, the second term of the RHS is weakly increasing in $x$.
There are two cases. First, $\psi^{\prime}(1)\geq0$, in which case $\forall
x\geq1,\psi^{\prime}(x)\geq0$, thus $\psi$ is weakly increasing and $\forall
x\geq1,\psi(x)\geq0$. Otherwise $\psi^{\prime}(1)<0$, in which case $%
\psi^{\prime}$ is first negative then positive above $1$ (since $%
\psi^{\prime}(\infty)=\infty$), hence $\psi$ is first decreasing and then
increasing, which also means that $\psi$ is first negative and then positive
above $1$. Therefore, $F_{21}$ FOSD $F_{22}$ if and only if $%
\psi^{\prime}(1)\geq0$. The condition reduces to
\begin{equation*}
b\leq\frac{2p-1}{2p}.
\end{equation*}

For (iv), working under the original model, previous equations reduce to $%
\Pi_{t_{0}}^{t}(1,1)=nm_{r}/m_{s}[p(\frac{t}{t_{0}})^{m_{s}}+(1-p)(\frac{t}{%
t_{0}})^{bm_{s})}-1]$, $\Pi_{t_{0}}^{t}(2,1)=nm_{r}/m_{s}(1-p)[(\frac{t}{%
t_{0}})^{m_{s}}-(\frac{t}{t_{0}})^{(bm_{s}}]$ and $\Pi_{t_{0}}^{t}(1,1)+%
\Pi_{t_{0}}^{t}(2,1)=nm_{r}/m_{s}[(\frac{t}{t_{0}})^{m_{s}}-1]$, which does
not depend on $p$. This proves the result.
\hbox{\hskip3pt\vrule width4pt
height8pt depth1.5pt}

\bigskip{}

\noindent \textbf{Proof of Proposition \ref{Prop_fosd} (page \pageref%
{Prop_fosd}):} Observe that $b$ increases with $\gamma$. This means that $%
(\Pi_{t_{0}}^{t})^{\prime }(1,1)\geq\Pi_{t_{0}}^{t}(1,1)$ and $%
(\Pi_{t_{0}}^{t})^{\prime }(2,2)\geq\Pi_{t_{0}}^{t}(2,2)$ while $%
(\Pi_{t_{0}}^{t})^{\prime }(2,1)\leq\Pi_{t_{0}}^{t}(2,1)$ and $%
(\Pi_{t_{0}}^{t})^{\prime }(1,2)\leq\Pi_{t_{0}}^{t}(1,2)$. The result then
follows from Lemma \ref{fosd}.
\hbox{\hskip3pt\vrule width4pt height8pt
depth1.5pt}

\bigskip{}

\noindent \textbf{Proof of Proposition \ref{Prop5} (page \pageref{Prop5}):}
Substituting from equation (\ref{b1 gM}), we have
\begin{equation*}
IH_{1}(p)=\frac{(1-2\gamma)^{2}\sigma p(1-p)}{\sigma
p(1-p)+(1+(1-2p)^{2}\sigma)\gamma(1-\gamma)}
\end{equation*}
From this expression, it is easily verified that $IH_{1}(p)=IH_{1}(1-p)$ and
that $IH_{1}(0)=IH_{1}(1)=0$. The first derivative of $IH_{1}$ is
\begin{equation*}
\frac{\partial IH_{1}(p)}{\partial p}=\frac{(1-2p)\sigma(\sigma+1)(2%
\gamma-1)^{2}\gamma(1-\gamma)}{\left((\sigma+1)\gamma(1-\gamma)+(2%
\gamma-1)^{2}p\sigma(1-p)\right)^{2}},
\end{equation*}
which has the same sign as $1-2p$, proving that $IH_{1}$ is increasing below
$p=1/2$ and then decreasing. To show concavity, write the second derivative
as
\begin{equation*}
\frac{\partial^{2}IH_{1}(p)}{\partial p^{2}}=\frac{2\gamma(1-\gamma)(2%
\gamma-1)^{2}\sigma(\sigma+1)\ast(\sigma(3p^{2}-3p+1)-\gamma(1-\gamma)(3%
\sigma(2p-1)^{2}-1))}{-\sigma^{3}(\gamma(1-\gamma)((2p-1)^{2}+1)+p(1-p))^{3}}%
.
\end{equation*}
The denominator is negative, and the term in the numerator before the
asterisk is positive, so $IH_{1}$ is concave if and only if $%
\sigma(3p^{2}-3p+1)-\gamma(1-\gamma)(3\sigma(2p-1)^{2}-1)>0$. Dividing by $%
\sigma$ and rearranging, we must show that $\gamma(1-\gamma)(3(2p-1)^{2}-1/%
\sigma)+3p(1-p)<1$. $\gamma(1-\gamma)\leq1/4$ and $-1/\sigma\leq0$; using
these inequalities and collecting terms proves the result. %
\hbox{\hskip3pt\vrule width4pt height8pt depth1.5pt}

\bigskip{}

\noindent \textbf{Proof of Corollary \ref{Coroll_shift} (page \pageref%
{Coroll_shift}):} The relevant derivatives are
\begin{eqnarray*}
\frac{\partial IH_{1}}{\partial\sigma} & = & \frac{(1-2\gamma)^{2}\gamma(1-%
\gamma)p(1-p)}{(\sigma p(1-p)+\gamma(1-\gamma)(1+(1-2p)^{2})\sigma)^{2}} \\
\frac{\partial IH_{1}}{\partial\gamma} & = & \frac{(2\gamma-1)p(1-p)%
\sigma(1+\sigma)}{(\sigma p(1-p)+\gamma(1-\gamma)(1+(1-2p)^{2})\sigma)^{2}},
\end{eqnarray*}
both of which are easily verified as being positive.
\hbox{\hskip3pt\vrule
width4pt height8pt depth1.5pt}

\subsection{Proofs for Section \protect\ref{sec_model_biased_search}.}

\noindent \textbf{Proof of Proposition \ref{prop-g1} (page \pageref{prop-g1}%
):} For what concerns the weak integration property, see the Proof of
Proposition \ref{prop_1}. The only thing to change is the right--hand part
of equation (\ref{rtb3}) instead of the formula in (\ref{PI2}).

\bigskip{}

For long--run integration, consider the solution to the general model, with
biased search, as described by equation (\ref{rtb4}). We follow the same
procedure as in the proof of Proposition \ref{prop_lri}, since $\mathbf{B}%
_{s}\odot\mathbf{A}$ is still a Markov Matrix. We obtain
\begin{eqnarray}
\lim_{t\rightarrow\infty}\frac{\Pi_{t_{0}}^{t}}{\pi_{t_{0}}(t)} & = & \left(%
\mathbf{B}_{s}\odot\mathbf{B}_{r}\right)^{-1}\mathbf{B}_{r}\mathbf{Q}\left(%
\begin{array}{c}
\vec{v}(\mathbf{\mathbf{B}_{s}\odot\mathbf{A}})^{\prime } \\
\vdots \\
\vec{v}(\mathbf{\mathbf{B}_{s}\odot\mathbf{A}})^{\prime }%
\end{array}%
\right)\mathbf{Q}^{-1}\ \ .
\end{eqnarray}

In the long run a node of type $i$ born at time $t_{0}$ will receive a
number of in--links from nodes of type $j$ which is a fraction

\begin{eqnarray}
\left[\lim_{t\rightarrow\infty}\frac{\Pi_{t_{0}}^{t}}{\pi_{t_{0}}(t)}\right]%
_{ij} & = & \sum_{h=1}^{H}\left(\left[\left(\mathbf{B}_{s}\odot\mathbf{B}%
_{r}\right)^{-1}\mathbf{B}_{r}\right]_{jh}p(h)\frac{\vec{v}(\mathbf{B}%
_{s}\odot\mathbf{A})_{i}}{p(i)}\right)  \notag \\
& = & \sum_{h=1}^{H}\left(\sum_{k=1}^{H}\left(\left[\left(\mathbf{B}_{s}\odot%
\mathbf{B}_{r}\right)^{-1}\right]_{jk}[\mathbf{B}_{r}]_{kh}\right)p(h)\frac{%
\vec{v}(\mathbf{B}_{s}\odot\mathbf{A})_{i}}{p(i)}\right)  \notag \\
& = & \sum_{h=1}^{H}\left(\sum_{k=1}^{H}\left(\left[\left(\mathbf{B}_{s}\odot%
\mathbf{B}_{r}\right)^{-1}\right]_{jk}p(k)\frac{p(k,h)}{p(h)}\right)p(h)%
\frac{\vec{v}(\mathbf{B}_{s}\odot\mathbf{A})_{i}}{p(i)}\right)  \notag \\
& = & \left(\sum_{h=1}^{H}\sum_{k=1}^{H}\left(\left[\left(\mathbf{B}_{s}\odot%
\mathbf{B}_{r}\right)^{-1}\right]_{jk}p(k)p(k,h)\right)\right)\frac{\vec{v}(%
\mathbf{B}_{s}\odot\mathbf{A})_{i}}{p(i)}  \notag \\
& = & \left(\sum_{k=1}^{H}\left[\left(\mathbf{B}_{s}\odot\mathbf{B}%
_{r}\right)^{-1}\right]_{jk}p(k)\right)p(j)\frac{\vec{v}(\mathbf{B}_{s}\odot%
\mathbf{A})_{i}}{p(i)}\ \ .  \label{eq_i}
\end{eqnarray}

of the overall links that it would receive in a type--blind process, where
the last line comes from the fact that $\sum_{h=1}^{H}p(k,h)=1$. The second
term is still a constant for type $i$, but the first term is generically not
proportional to $p(j)$. \hbox{\hskip3pt\vrule width4pt height8pt depth1.5pt}

\bigskip{}

Finally, for the partial integration property, the proof is analogous to the
Proof of Proposition \ref{prop_partial_no_bias}. In this case
\begin{equation*}
\frac{\Pi_{t_{0}}^{t}}{\pi_{t_{0}}(t)}=\left(\mathbf{B}_{s}\odot\mathbf{B}%
_{r}\right)^{-1}\mathbf{B}_{r}\mathbf{Q}\left(\left(\left(\frac{t}{t_{0}}%
\right)^{m_{s}}-1\right)^{-1}\sum_{\mu=1}^{\infty}\frac{\left(m_{s}\log\left(%
\frac{t}{t_{0}}\right)\right)^{\mu}}{\mu!}(\mathbf{B}_{s}\odot\mathbf{A}%
)^{\mu}\right)\mathbf{Q}^{-1}\ \ .
\end{equation*}
As $\mathbf{B}_{s}\odot\mathbf{A}$ satisfies the monotone convergence
property, we can use Lemma \ref{matrix_monotone} from \ref{matrices}, and
the same reasoning applies.
\hbox{\hskip3pt\vrule width4pt height8pt
depth1.5pt}

\pagebreak{}

\end{document}